\documentclass[10pt,tightenlines,prl,showpacs,letterpaper,aps,twocolumn,nofootinbib,superscriptaddress]{revtex4-1}

\usepackage{graphicx}
\usepackage{amssymb}
\usepackage{epstopdf}
\usepackage{amsmath,amsthm}
\usepackage[normalem]{ulem} 
\usepackage{comment}
\usepackage{wasysym}
\usepackage{subfigure}
\usepackage{hyperref}
\usepackage{placeins}

\newcommand{\COMMENT}[1]{}

\newcommand{\beq}{\begin{equation}}
\newcommand{\eeq}{\end{equation}}
\newcommand{\bea}{\begin{align}}
\newcommand{\eea}{\end{align}}
\newcommand{\bq}{\begin{quote}}
\newcommand{\eq}{\end{quote}}

\newcommand{\blk}{\color{black}}

\usepackage{xparse}

\usepackage[x11names]{xcolor}
\ExplSyntaxOn
\newcommand{\colorstring}[3]{%
	\str_set:Nn \l_tmpa_str {#3}
	\int_step_inline:nnnn {1} {1} {\str_count:N \l_tmpa_str } {%
		\int_if_odd:nTF{##1}{
			\textcolor{#1}{\str_item:Nn \l_tmpa_str {##1}}
		}{
			\textcolor{#2}{\str_item:Nn \l_tmpa_str {##1}}
		}%
	}%
}
\ExplSyntaxOff
\begin{document}

\title{What is nonclassical about uncertainty relations?}
\author{Lorenzo Catani}\email{lorenzo.catani@tu-berlin.de}\affiliation{Electrical Engineering and Computer Science Department, Technische Universit\"{a}t Berlin, 10587 Berlin, Germany}

\author{Matthew Leifer}
\affiliation{Institute for Quantum Studies and Schmid College of Science and Technology, Chapman University, One University Drive, Orange, CA, 92866, USA}

\author{Giovanni Scala}
\affiliation{International Centre for Theory of Quantum Technologies, University of Gdansk, 80-308 Gdansk, Poland}

\author{David Schmid}
\affiliation{International Centre for Theory of Quantum Technologies, University of Gdansk, 80-308 Gdansk, Poland}

\author{Robert W. Spekkens}
\affiliation{Perimeter Institute for Theoretical Physics, 31 Caroline Street North, Waterloo, Ontario Canada N2L 2Y5}

\begin{abstract}
Uncertainty relations express limits on the extent to which the outcomes of distinct measurements on a single state can be made jointly predictable.  The existence of nontrivial uncertainty relations in quantum theory 
 is generally considered to be a way in which it entails
 a departure from the classical worldview. However, this perspective is undermined 
  by the fact that there exist operational theories which exhibit nontrivial uncertainty relations but which are consistent with the classical worldview insofar as they admit of 
  a generalized-noncontextual ontological model.
This prompts the question of what 
 aspects of uncertainty relations, if any, {\em cannot} be realized in this way
 and so constitute evidence of genuine nonclassicality.
We here consider uncertainty relations describing the tradeoff between the predictability of a pair of binary-outcome measurements (e.g., measurements of Pauli $X$ and Pauli $Z$ observables in quantum theory).
We show that, for a class of theories satisfying a particular symmetry property, the functional form of this predictability tradeoff is constrained by noncontextuality to be below a linear curve.  
Because qubit quantum theory has the relevant symmetry property, the fact that its predictability tradeoff describes a section of a circle is a violation of this noncontextual bound, and therefore constitutes an example of how the functional form of an uncertainty relation can witness contextuality.  We also deduce the implications for a selected group of operational foils to quantum theory and consider the generalization to three measurements. 
 \blk
\end{abstract}

\maketitle

A wide range of phenomena have been viewed as intrinsically quantum, in the sense that they are thought to resist classical explanation---noncommutativity, interference, collapse, no-cloning, teleportation, remote steering, and entanglement, to name just a few. However, the aspects of all of these phenomena (and many more) that have traditionally been regarded as relevant to establishing this claim
 can in fact be reproduced in a noncontextual\footnote{In this article, we will use the term ``noncontextual'' to refer to the notion of generalized noncontextuality, defined in Ref.~\cite{Spekkens2005}, which is distinct from the Kochen-Specker notion of noncontextuality~\cite{KochenSpecker1967} and reduces to the latter in certain circumstances.} ontological model~\cite{Spekkens2005}, as demonstrated in Refs.~\cite{Spekkens2007,Bartlett2012,chiribella2016quantum,ToyFieldTheory}.  Therefore, if one takes the possibility of a noncontextual ontological model as a good notion of classical explainability (there are many arguments in favour of doing so; see Section~V.A.3 of Ref.~\cite{ToyFieldTheory} or the introduction of Ref.~\cite{schmid2021guiding}), then the possibility of reproducing 
 these aspects 
 undermines the 
claim that they   resist classical explanation. 
This prompts the question: for each item on the list, are there more nuanced aspects of the full phenomenology that actually {\em do} resist explanation in terms of a noncontextual ontological model? In other words: what is genuinely nonclassical about its phenomenology?  This question has been investigated, for instance, for minimum-error state discrimination \cite{SchmidSpekkens2018}, unambiguous state discrimination~\cite{flatt2021,roch2021}, state-dependent cloning \cite{lostaglio2020contextual}, scenarios with pre- and post-selection~\cite{
pusey2015logical,Pusey2014,KunjwalLostaglioPusey2019}, and linear response theory~\cite{lostaglio2020certifying}.  
 
This letter undertakes an investigation of what is genuinely nonclassical about uncertainty relations.  Many different notions have been termed ``uncertainty relations''.  We are here concerned with the version that asserts that
there are pairs of measurements for which there is a nontrivial tradeoff in their predictabilities.\footnote{This is distinct from the version that asserts that any information gain about a system necessarily disturbs it, often termed an error-disturbance relation~\cite{busch2014colloquium} 
  and the version  that asserts a tradeoff in the degree of noise one must add to make incompatible observables compatible~\cite{busch2007approximate,bullock2018measurement}, termed an error tradeoff relation, which is related to the error-disturbance relation.  }  Previous works have noted that there are operational theories that admit of a noncontextual ontological model  and for which an uncertainty relation holds, such as Gaussian quantum mechanics~\cite{Bartlett2012} and the stabilizer theory of qudits where $d$ is an odd prime~\cite{chiribella2010probabilistic,Catani2017}.
Thus, although it is conventionally thought that the mere existence of an uncertainty relation is an intrinsically quantum phenomenon, the fact that this happens in theories that admit of a noncontextual ontological model demonstrates that it is not at odds with the classical worldview.   The question, therefore, is whether one can identify {\em other aspects} of uncertainty relations
 that provably {\em cannot} arise in a noncontextual ontological model.  

We here demonstrate that for a certain class of operational theories, an uncertainty relation describing the predictability tradeoff for a pair of binary-outcome measurements can witness contextuality through its functional form. 

{\em A class of uncertainty relations for a qubit.}
In the early days of quantum theory, uncertainty relations were formulated in terms of products of standard deviations~\cite{Heisenberg1925,Kennard1927,Robertson1929}.  
As has been pointed out by several authors~\cite{Deutsch1983,Maccone2014}, these are unsatisfactory for finite-dimensional systems because they involve a bound that depends on the state and which can be trivial for certain states.  One solution to this problem is to focus on {\em sums} of standard deviations rather than {\em products}, because such sums satisfy a nontrivial bound for {\em all} states. 
We will here focus on the Pauli $X$ and $Z$ observables, which are complementary and represent discrete analogues of position and momentum.
The strongest uncertainty relation that can be derived for $X$ and $Z$ is $\Delta X^2 + \Delta Z^2 \ge 1$,
as we demonstrate in Appendix~\ref{Appendix_UR}.  There, we show that this can be written in several other useful forms,
one of which is the form in which it was (to our knowledge) first proposed~\cite{Luis2003}, by building on the work of Ref.~\cite{Larsen1990}.  The form that we will prefer 
for the purposes of this article is 
\begin{align}\label{QUR0}
\langle X\rangle^2 + \langle Z\rangle^2 \le 1.
\end{align} 
We will be taking our preferred measure of predictability to be the absolute values of the expectation values, i.e., $|\langle X\rangle|$ and $|\langle Z\rangle|$, so that Eq.~\eqref{QUR0} expresses a tradeoff relation between the squares of these predictabilities for every state.  It is therefore apt to refer to Eq.~\eqref{QUR0} as the quantum $\mathrm{ZX}$-uncertainty relation.

Note that Eq.~\eqref{QUR0} follows  trivially from 
\begin{equation}\label{quantumXYZUR}
\langle X\rangle^2 + \langle Y\rangle^2 + \langle Z\rangle^2 \le 1,
\end{equation}
a relation we refer to as the quantum {\em $\mathrm{XYZ}$-uncertainty relation} and whose validity follows from the fact that it is a description of the Bloch ball of qubit quantum states.

{\em Operational theories.}
 In prepare-measure scenarios, to which we limit ourselves here, an operational theory stipulates the possible preparations of a system and the possible measurements thereon, as well as an algorithm for computing the probability $\mathbb{P}(y|M,P)$ of obtaining the outcome $y$ of measurement $M$ given preparation $P$, for all possible measurements and preparations. For the purposes of making predictions, it is possible to represent each preparation $P$ and each effect $[y|M]$ by real-valued vectors $\vec{s}_P$ and $\vec{e}_{y|M}$ respectively, with $\mathbb{P}(y|M,P)= \vec{s}_P \cdot \vec{e}_{y|M}$~\cite{hardy2001quantum, barrett2007information,chiribella2010probabilistic}. 

Quantum theory can be conceptualized as an operational theory, but one can also consider operational theories that make different predictions.  These are typically studied because of what they can teach us about quantum theory via the {\em contrast} they provide with it.  For this reason, they are termed {\em foil theories}~\cite{chiribella2016quantum}.

\begin{figure*}[hbt!]
	\centering
   \subfigure[$\,$ qubit theory]{\includegraphics[width=1.38in]{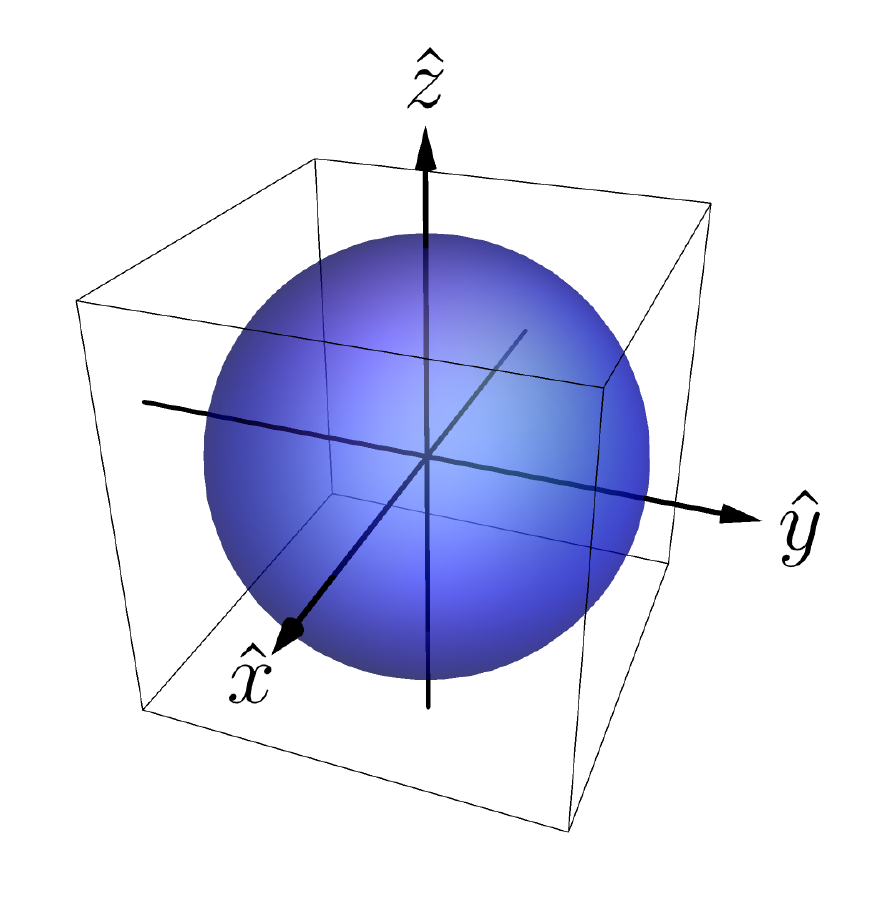}}
   \subfigure[$\,$ stabilizer theory]{\includegraphics[width=1.38in]{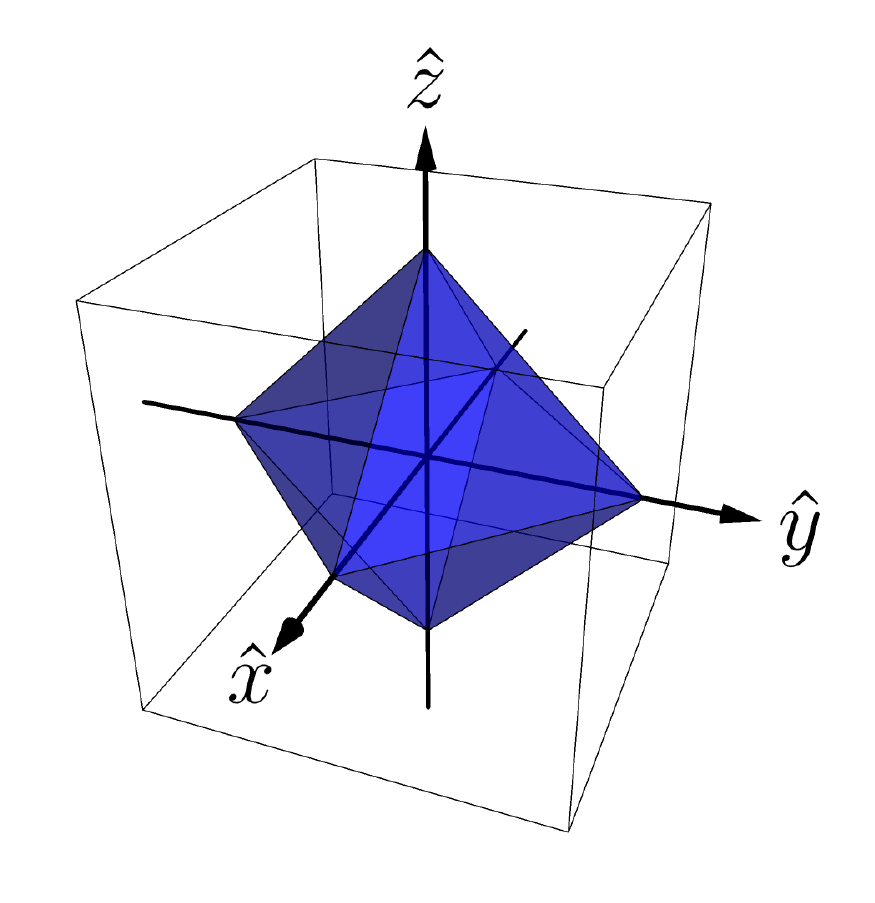}}
   \subfigure[$\,\eta$-depolarized qubit theory]{\includegraphics[width=1.38in]{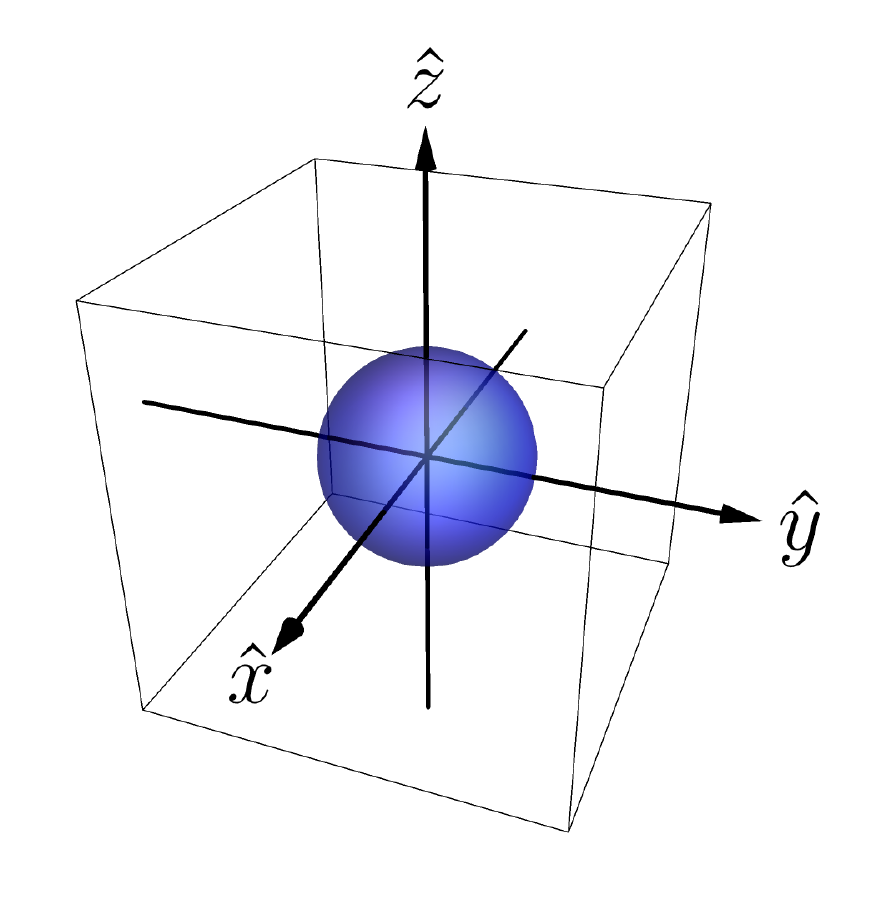}}
   \subfigure[$\,$ gbit theory]{\includegraphics[width=1.38in]{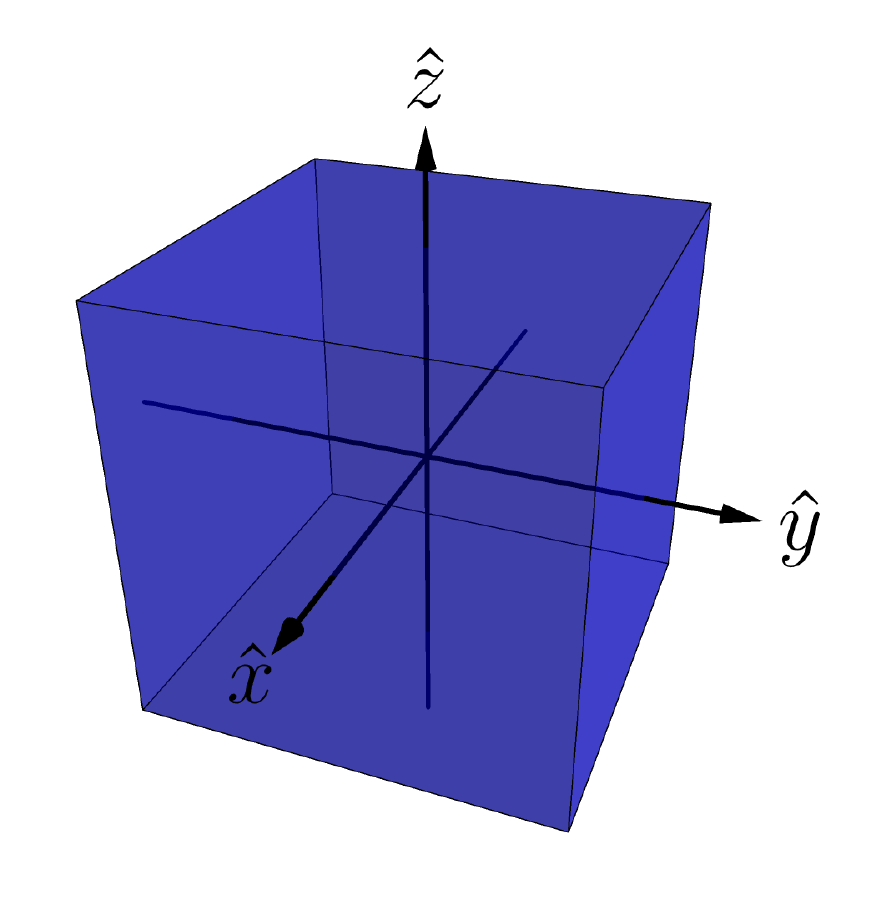}}
   \subfigure[$\,$ simplicial theory]{\includegraphics[width=1.38in]{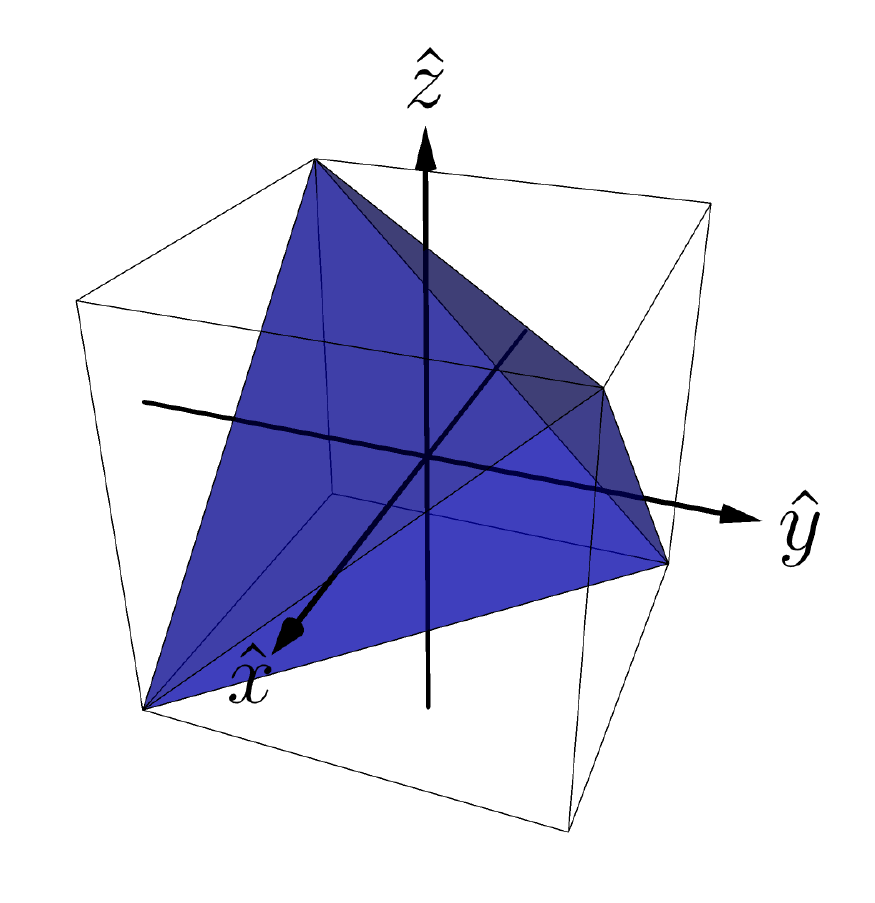}}
   \caption{The state spaces of various operational theories.
    }
   \label{fragments}
\end{figure*}

We discuss four examples of operational foils to qubit quantum theory that  provide a useful contrast in the domain of uncertainty relations and that have been of independent prior interest (see Fig.~\ref{fragments}). 

Because the real-valued vector representation of qubit quantum theory is simply the familiar 4-dimensional Bloch representation (wherein every qubit operator is represented as a linear combination of elements of a basis of the 4-dimensional space of Hermitian operators), we consider foil theories that also have a 4-dimensional real-valued vector representation. In discussing these theories, we will reuse the notation $X$, $Y$, and $Z$ to refer to a triple of measurements associated with directions in the real-valued representation that are mutually orthogonal to one another and to the unit effect.

The first two foil theories are {\em subtheories} of the qubit theory, in the sense that they posit that only a {\em subset} of the preparations and measurements thereof are physically possible.  First is the {\em qubit stabilizer theory}, defined as the subtheory of the full qubit theory arising when the states are restricted to the convex hull of the stabilizer states (an octahedron embedded inside the Bloch sphere) and the effects are restricted to the closure (under both convex mixtures and coarse-grainings) of stabilizer effects. It has been of prior interest in quantum information theory~\cite{Gottesman1997} and quantum foundations~\cite{chiribella2016quantum,Catani2017}.  
Second is the {\em $\eta$-depolarized qubit theory}, defined by taking the set of effects to be the full set of qubit effects, but taking the states to be restricted to the image of the Bloch ball under the $\eta$-depolarizing map $\mathcal{D}_{\eta}(\rho) \equiv (1-\eta) \rho + \eta \frac{1}{2}I$.  The state space in this case corresponds to  a contracted Bloch ball of radius $1-\eta$. Note that because this is being considered as a foil theory, the  depolarization is imagined to be fundamental, i.e., the theory is stipulated to have intrinsic decoherence of the type explored in collapse theories~\cite{milburn1991intrinsic,ghirardi1986unified}.
Our third example of a foil theory is one that is {\em post-quantum}, in the sense that it predicts statistics in prepare-measure scenario that are {\em not} achievable in quantum theory. This is the {\em gbit theory}~\cite{barrett2007information,short2010strong}, but defined relative to {\em three} binary-outcome measurements  rather than a pair.  We will refer to these three measurements as $X$, $Y$, and $Z$, in analogy with the quantum case.   
As such, we can describe the states and effects of the gbit theory in the same real vector space as we used for the other foil theories: the effect space of the gbit theory is equivalent to that of the qubit stabilizer theory,
 while the state space is a cube~\cite{mazurek2021experimentally}. The gbit theory has been studied extensively in the context of axiomatizing quantum theory~\cite{Popescu1994,krumm2017thermodynamics}. 
Finally, our fourth example of a foil theory is a strictly classical theory describing a {\em pair} of binary random variables, which we again denote by $X$ and $Z$.  This {\em simplicial theory} 
 has a state space which is the convex hull of the four possible joint assignments of values to $X$ and $Z$, which is a regular tetrahedron, while the effect space is the 4-dimensional hypercube that is dual to this simplex~\cite{barrett2007information,janotta2014}.

\begin{figure}[htb!]
   \centering
   \includegraphics[width=\columnwidth]{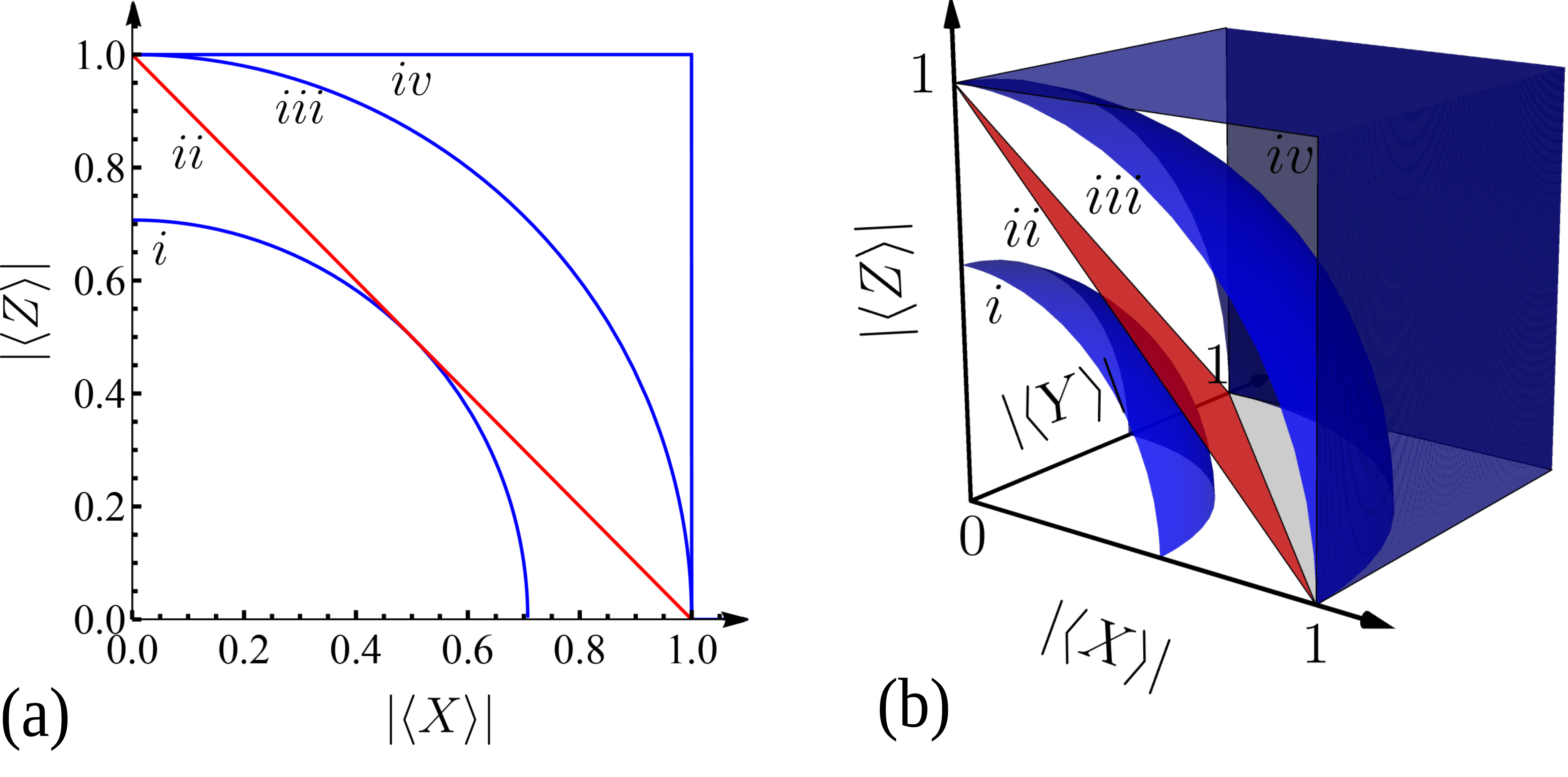}
   \caption{ (a) The $\mathrm{ZX}$-uncertainty relation for (i) $\eta$-depolarized qubit theory for $\eta=1-\frac{1}{\sqrt{2}}$, (ii) stabilizer qubit theory, (iii) qubit theory, and (iv) gbit theory and simplicial theory. Curve (ii) also describes the noncontextual bound.  (b) The $\mathrm{XYZ}$-uncertainty relation for (i) $\eta$-depolarized qubit theory for $\eta=1-\frac{1}{\sqrt{3}}$, (ii) stabilizer qubit theory, (iii) qubit theory, and (iv) gbit theory and simplicial theory. Surface (ii) also describes the noncontextual bound. 
   }
   \label{Tradeoffs}
\end{figure}

The $\mathrm{ZX}$-uncertainty relations of the four foil theories described above are as follows:
\begin{align}
\textrm{qubit stabilizer:}& \;\; |\langle X\rangle| + |\langle Z\rangle| \le 1,\label{stabUR}\\
\textrm{$\eta$-depolarized qubit:}&\;\; \langle X\rangle^2 + \langle Z\rangle^2 \le (1- \eta)^2, \label{etaUR}\\
\textrm{gbit:}& \;\; |\langle X\rangle| \le 1, |\langle Z\rangle| \le 1, \label{gbitUR}\\
 \textrm{simplicial:}& \;\; |\langle X\rangle| \le 1, |\langle Z\rangle| \le 1.\label{simplicialUR}
\end{align}
These are  implied by the geometry of the projection of their state spaces into the $\hat{x}\hat{z}$-plane, and are plotted in Fig.~\ref{Tradeoffs}(a) alongside the quantum $\mathrm{ZX}$-uncertainty relation. Note that the relations for the gbit and simplicial theory describe a {\em lack} of any nontrivial tradeoff, i.e., both $X$ and $Z$ can be made perfectly predictable simultaneously. 

{\em Ontological models and noncontextuality.} An ontological model of an operational theory 
 is defined as follows. 
 For each system, the model specifies a set $\Lambda$, termed an ontic state space, describing the possible physical states, or {\em ontic states}, of the system, denoted $\lambda \in \Lambda$. (For our purposes, it suffices to consider $\Lambda$ finite.)
 Each preparation procedure $P$ in the operational theory is represented as a probability distribution over the ontic states, denoted $\mu(\lambda |P)$. For each measurement $M$ and outcome $y$ of $M$, the effect $[y|M]$ is 
  represented by a conditional probability distribution, denoted  $\xi(y|M,\lambda)$, that stipulates the probability of obtaining outcome $y$  given that the measurement $M$ was implemented on the system and that the latter was in the ontic state $\lambda$.  It is often useful to view the probability distribution $\mu(\lambda|P)$ as a vector 
  denoted $\vec{\mu}_P$, and to also view the conditional probability distribution $\xi(y|M,\lambda)$ as a vector
   denoted $\vec{\xi}_{y|M}$.  
 It follows that the model reproduces the predictions of the operational theory if and only if
\begin{align} \label{opdata}
\mathbb{P}(y|M,P)
&=\sum_{\lambda\in\Lambda} \xi(y|M,\lambda)\mu(\lambda |P)= \vec{\xi}_{y|M} \cdot \vec{\mu}_P.
\end{align}

The principle of generalized noncontextuality, applied to preparation procedures\footnote{Here, we will not need to apply the principle of noncontextuality to other sorts of experimental procedures, e.g., measurements or transformations.}, has the following form: 
two preparation procedures, $P$ and $P'$, that are {\em operationally equivalent} (defined as leading to the same statistics for all possible measurements, $\forall M: \mathbb{P}(y|M,P)= \mathbb{P}(y|M,P')$, and denoted $P\simeq P'$) must be represented in the ontological model by the same probability distribution over ontic states:
\begin{equation}
P\simeq P' \implies \mu(\lambda |P)=\mu(\lambda |P').
\end{equation}

The real-valued vector representation $\vec{s}_P$ of a preparation $P$, described earlier, throws away all information about $P$ besides its operational equivalence class.  It follows that a noncontextual ontological model of an operational theory is one wherein all preparation procedures associated to the same vector $\vec{s}$ are represented  by the same probability distribution over ontic states.  In particular, this implies that if two different mixtures of operational states are equal, the same relation holds among the corresponding probability distributions over ontic states:
\begin{align} \label{linearnc}
\sum_i w_i \vec{s}_{i} = \sum_j w'_j \vec{s}^{\;\prime}_{j}\;\;\implies\;\; \sum_i  w_i \vec{\mu}_i = \sum_j w'_j \vec{\mu}^{\;\prime}_{j}
\end{align}
where $\{w_i\}_i$ and $\{w'_{j}\}_{j}$ are probability distributions~\cite{Spekkens2005}.\blk

Quantum theory, conceived as an operational theory,  does not admit of a 
 preparation-noncontextual ontological model even for a single qubit~\cite{Spekkens2005}.
By contrast, the qubit stabilizer theory, when restricted to a single system in a prepare-measure scenario, admits of a noncontextual ontological model~\cite{Spekkens2007}.   
The $\eta$-depolarized qubit theory admits of a noncontextual model for $\eta \ge \frac{2}{3}$~\cite{marvian2020inaccessible}. 
\blk 
 The gbit theory, like the qubit quantum theory, does not admit of a noncontextual ontological model.
   Finally, the simplicial theory admits of a noncontextual ontological model where the vertices of the simplex are themselves the ontic states~\cite{schmid2021simplexembeddability}.

{\em Main result.} 
There is an immediate challenge with trying to cast an uncertainty relation as a noncontextuality inequality. An uncertainty relation expresses a predictability tradeoff between two measurements {\em for any single quantum state}. But the simplest operational scenario in which noncontextuality implies a nontrivial constraint on statistics involves {\em four} quantum states~\cite{pusey2018robust}, since this is the smallest number for which there can be a nontrivial operational equivalence.

 To see how one solves this problem, consider the case of the qubit theory.
Note that for any given quantum state, the values of $\mathrm{X}$-predictability and $\mathrm{Z}$-predictability that it achieves can {\em also} be achieved by many other states, and one can find nontrivial operational equivalences among these.
In particular, imagine a state with Bloch vector $\vec{s}_1$. Then one can find three other states $\vec{s}_2$, $\vec{s}_3$, and $\vec{s}_4$ that give the same predictabilities, but different signs for the expectation values; that is,
\begin{align}\label{eq:rectanglesymmetryEV}
& \langle X\rangle_{\vec{s}_1}= - \langle X\rangle_{\vec{s}_2} = - \langle X\rangle_{\vec{s}_3} = \langle X\rangle_{\vec{s}_4},\nonumber\\
&\langle Z\rangle_{\vec{s}_1} = \langle Z\rangle_{\vec{s}_2} = - \langle Z\rangle_{\vec{s}_3} = - \langle Z\rangle_{\vec{s}_4}.
\end{align}
We refer to this as the condition that the state has {\em equal predictability counterparts}. 
Moreover, one can always find such quadruples of states which additionally satisfy the operational equivalence relation
\begin{equation}\label{eq:op_equiv}
\frac{1}{2}\vec{s}_1+ \frac{1}{2}\vec{s}_3 = \frac{1}{2}\vec{s}_2+ \frac{1}{2}\vec{s}_4. 
\end{equation}

\begin{figure}[htb!] 
	\centering
        \includegraphics[width=1.8in]{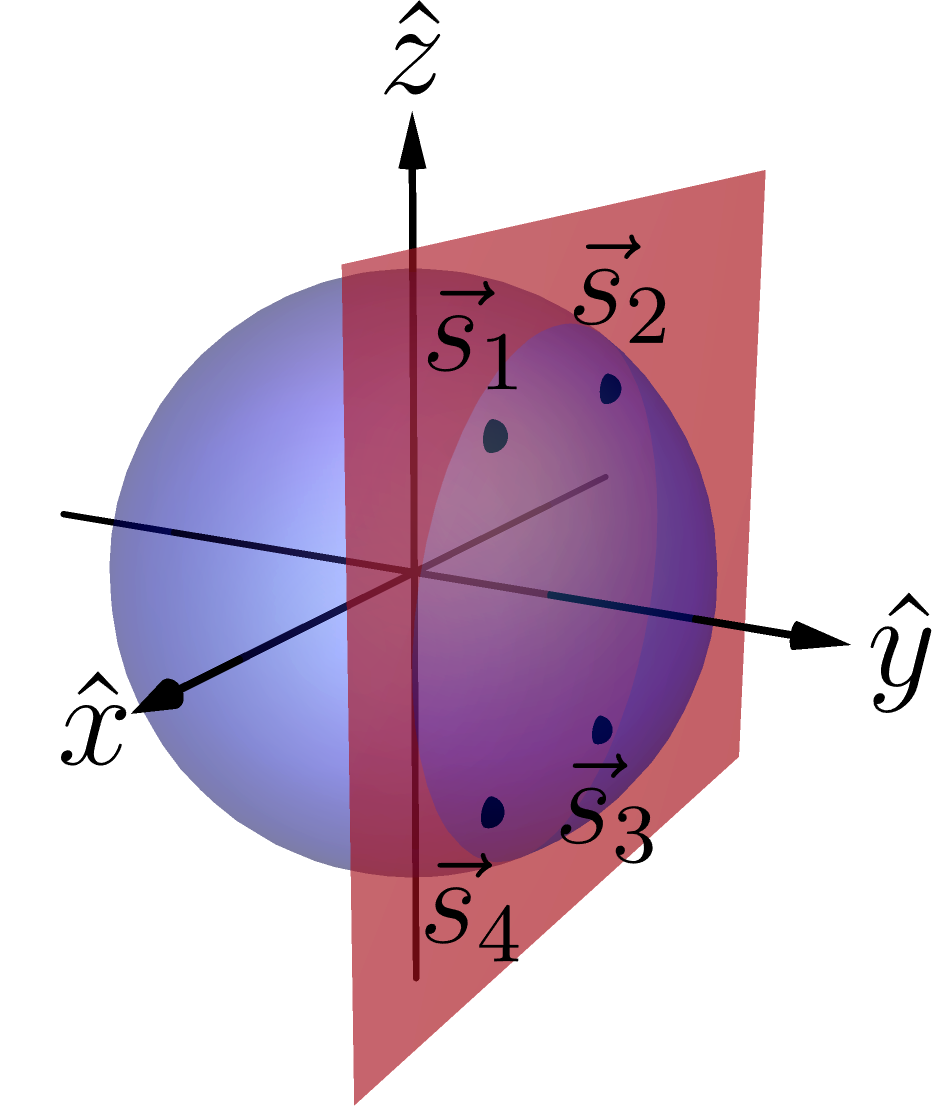}
	\caption{Depiction of how an arbitrary state $\vec{s}_1$ in qubit quantum theory is part of a quadruple of states that satisfy the $A_1^2$-orbit-realizability condition. }
	\label{fig:statesmmts}
\end{figure}

An example is depicted in Fig.~\ref{fig:statesmmts}. Such a quadruple of states forms the vertices of a rectangle in a plane that is parallel to the $\hat{x}\hat{z}$-plane.   These vertices are the orbit of the original state under the action of the symmetry group of a rectangle under reflections, the Coxeter group $A_1^2$,
 so we refer to the pair of conditions on the state as the condition of {\em $A_1^2$-orbit-realizability}.

 Our main technical result is that, in any operational theory, if one can find a pair of measurements, which we will here denote by $X$ and $Z$, and a state that satisfies the $A_1^2$-orbit-realizability condition (where the $A_1^2$ symmetry is evaluated relative to $X$ and $Z$), then noncontextuality implies a nontrivial constraint on the $\mathrm{X}$-predictability and $\mathrm{Z}$-predictability for that state, namely, that they satisfy
\begin{align}\label{NCURgeneral}
|\langle X\rangle| + |\langle Z\rangle| \le 1.
\end{align}
An analytic and self-contained proof of this claim is given in Appendix~\ref{appendix_MainProof}.  In Appendix~\ref{appendix_AlternativeProof}, we show that it also follows as a special case of noncontextuality inequalities that were previously derived using a linear program~\cite{schmid2018all}. Note that the noncontextuality inequality of Eq.~\eqref{NCURgeneral} is noise-robust, and can therefore be tested experimentally using the techniques described in Refs.~\cite{mazurek2016experimental} and~\cite{mazurek2021experimentally}. 

Eq.~\eqref{NCURgeneral} has an unconventional form for a noncontextuality inequality given that it constrains the predictions associated to a {\em single} state rather than a set of states.
This difference is only cosmetic, however, as the  single state is explicitly required to satisfy the $A_1^2$-orbit-realizability condition and thus the predictabilities appearing in the inequality in fact refer to the data one can obtain from {\em any} of the quadruple of states in its $A_1^2$-orbit.\blk

For any operational theory and choice of $X$ and $Z$ measurements in that theory, one can determine the subset of states that satisfy the $A_1^2$-orbit-realizability condition relative to that choice.  In the case of the simplicial theory, depicted in Fig.~\ref{fragments}(e), for instance, it is the strict subset of states defined by the octahedron whose vertices lie at the midpoints of the edges of the tetrahedron  (i.e., the octahedron depicted in Fig.~\ref{fragments}(b)). 
 A vertex of the tetrahedron, for example, fails to satisfy the $A_1^2$-orbit-realizability condition because although it satisfies the condition of having equal predictability counterparts (namely, the three other vertices), these four states 
    do not satisfy the operational equivalence condition.
       By contrast,  there are operational theories wherein {\em all} states satisfy the  $A_1^2$-orbit-realizability condition.  Examples include the qubit theory, the stabilizer qubit theory, the $\eta$-depolarized qubit theory, and the gbit theory.  We will refer to operational theories of this sort as having $A_1^2$-symmetry. 

Whether or not an operational theory has $A_1^2$-symmetry,
  our bound constrains the tradeoff between $\mathrm{X}$-predictability and $\mathrm{Z}$-predictability for any state within the theory that satisfies the $A_1^2$-orbit-realizability condition.  Consequently, if the theory contains one or more such states that {\em violate} the inequality, this is a proof of the failure of that theory to admit of a noncontextual ontological model. 

For operational theories that {\em do} have $A_1^2$-symmetry,
our bound has further significance.  Because in such theories {\em all states} satisfy the $A_1^2$-orbit-realizability condition, our bound is a universal constraint on the predictability tradeoff within such theories, that is, it is a constraint on
{\em the form of the $\mathrm{ZX}$-uncertainty relation} within such theories.

The noncontextual bound (Eq.~\eqref{NCURgeneral}) is compared to the $\mathrm{ZX}$-uncertainty relation for a qubit (Eq.~\eqref{QUR0}) in Fig.~\ref{Tradeoffs}(a), where it is readily seen that there can be quantum violations of the bound.   Indeed, only when $|\langle X\rangle|=1$ or $|\langle Z\rangle|=1$ does the noncontextual bound intersect the quantum $\mathrm{ZX}$-uncertainty relation. 
The maximum quantum violation 
is achieved when $|\langle X\rangle|= |\langle Z\rangle|= \frac{1}{\sqrt{2}}$ and corresponds to 
$|\langle X\rangle| + |\langle Z\rangle|= \sqrt{2} \simeq 1.414$.

One can also compare this
noncontextual bound with the $\mathrm{ZX}$-uncertainty relation of the three foil theories that are in the $A_1^2$-symmetry class.  The $\mathrm{ZX}$-uncertainty relation for the $\eta$-depolarized qubit theory, Eq.~\eqref{etaUR}, satisfies the noncontextual bound if $\eta \ge 1-\frac{1}{\sqrt{2}}\simeq 0.293$.
  The $\mathrm{ZX}$-uncertainty relation for the qubit stabilizer theory, Eq.~\eqref{stabUR}, has exactly the same form as 
Eq.~\eqref{NCURgeneral} and therefore precisely saturates the noncontextual bound.   Finally, the uncertainty relation for the gbit theory,  Eq.~\eqref{gbitUR}, yields the maximum possible violation of the noncontextual bound, namely, $|\langle X \rangle| + |\langle Z \rangle| =2$.

By contrast, because the simplicial theory is {\em not} in the $A_1^2$-symmetry class, our result does not constrain the form of its $\mathrm{ZX}$-uncertainty relation. Therefore, although the $\mathrm{ZX}$-uncertainty relation for the simplicial theory, Eq.~\eqref{simplicialUR}, is equivalent to that of the gbit theory and thus can violate the bound of Eq.~\eqref{NCURgeneral}, the only states in the theory that achieve this violation (for example, the vertices of the simplex) do not satisfy $A_1^2$-orbit realizability and 
Eq.~\eqref{NCURgeneral} is not derivable from noncontextuality for them. 
Meanwhile, the states that {\em do} satisfy the $A_1^2$-orbit realizability condition are precisely those inside of the embedded octahedron, namely, the states arising in the qubit stabilizer theory, and these saturate the noncontextual bound.
 In short, contextuality is not witnessed in the case of the simplicial theory, consistent with the fact that the latter admits of a noncontextual model.

{\em Generalization to three measurements.} 
The analogue of Eq.~\eqref{NCURgeneral} for {\em three} measurements (which we denote $X$, $Y$, and $Z$) is
\begin{equation}\label{NCboundXYZ}
|\langle X\rangle| + |\langle Y\rangle| + |\langle Z\rangle| \le 1.
\end{equation}
In Appendix~\ref{Appendix_3Dcase}, we articulate the condition of $A_1^3$-orbit-realizability under which this bound holds ($A_1^3$ is the symmetry group of a rectangular prism under reflections) and provide the proof. 
This constraint is depicted in red in Fig.~\ref{Tradeoffs}(b), alongside the $\mathrm{XYZ}$-uncertainty relations for the four foil theories discussed above.  Note that this inequality admits of a greater quantum violation than Eq.~\eqref{NCURgeneral} does.  The stabilizer qubit theory also saturates this inequality. 


{\em Discussion.}
It is usually the {\em lack} of joint predictability of $X$ and $Z$ (or of $X$, $Y$ and $Z$) that is emphasized as a feature of quantum theory that constitutes a departure from the classical worldview.  From this perspective, what is striking about our results is that qubit quantum theory contains states that assign {\em higher} values of the predictabilities of multiple measurements, such as $X$ and $Z$ (or $X$, $Y$ and $Z$) than can occur in any operational theory that is  noncontextually realizable (hence classically explainable) and that has $A_1^2$-symmetry.   Similarly, the fact that the gbit theory can achieve {\em perfect} predictability for $X$ and $Z$ jointly (and even for $X$, $Y$ and $Z$ jointly) while having $A_1^2$-symmetry implies that it is even {\em further} than the qubit theory from being classically explainable.  

The $A_1^2$-symmetry property is critical to understanding why the degree of nonclassicality increases with the degree of predictability rather than with the degree of {\em un}predictability.  The conventional association of nonclassicality with unpredictability is based  on the fact that the simplicial theory---which must surely be included among those that are classically explainable---allows perfect joint predictability of $X$ and $Z$.  However, the states in the simplicial theory that achieve such predictability do not satisfy the $A_1^2$-orbit-realizability condition, and hence their ontological representations are not constrained by noncontextuality. Moreover, as noted above, if one considers the subset of states within the simplicial theory that {\em do} satisfy the $A_1^2$-orbit-realizability condition (namely, the embedded octahedron), they exhibit {\em less} joint predictability for $X$ and $Z$ than is possible in qubit quantum theory.
{\em Acknowledgments}---We thank Elie Wolfe for helping us to verify our noncontextual bound for three measurements.
This research was supported by Perimeter Institute for Theoretical Physics and by QuantERA/2/2020, an ERA-Net co-fund in Quantum Technologies, under the eDICT project.
 Research at Perimeter Institute is supported by the Government of Canada through the Department of Innovation, Science and Economic Development Canada and by the Province of Ontario through the Ministry of Research, Innovation and Science.  RWS was also supported by the Natural Sciences and Engineering Research Council of Canada (Grant No. RGPIN-2017-04383). 
ML was supported in part by the Fetzer
Franklin Fund of the John E. Fetzer Memorial Trust and
by grant number FQXi-RFP-IPW-1905 from the Foundational Questions Institute and Fetzer Franklin Fund, a donor advised fund of Silicon Alley Community Foundation.
LC acknowledges funding from the Einstein Research Unit `Perspectives of a Quantum Digital Transformation'.
DS and GS acknowledge support by the Foundation for Polish Science (IRAP project, ICTQT, contract no.2018/MAB/5, co-financed by EU within Smart Growth Operational Programme). 

\bibliographystyle{apsrev4-1}
\bibliography{bib}



\appendix

\section{Proof of the noncontextual bound}
\label{appendix_MainProof}


Consider an operational theory and a pair of measurements, denoted $X$ and $Z$.  Denote the real-valued vector representing outcome $\pm1$ of measurement $X$ (respectively $Z$) by $\vec{e}_{\pm1|X}$ (respectively $\vec{e}_{\pm1|Z}$). For a preparation represented by the real-valued vector $\vec{s}$, the expectation values of $X$ and $Z$ are defined as
\begin{align}
\langle X \rangle_{\vec{s}} &= (\vec{e}_{+1|X} - \vec{e}_{-1|X})\cdot \vec{s},\nonumber\\
\langle Z \rangle_{\vec{s}} &= (\vec{e}_{+1|Z} - \vec{e}_{-1|Z})\cdot \vec{s}.
\end{align} 
The $\mathrm{X}$-predictability and $\mathrm{Z}$-predictability for $\vec{s}$ are then defined as $|\langle X \rangle_{\vec{s}} |$ and $|\langle Z \rangle_{\vec{s}} |$.

Now consider a state $\vec{s}_1$ whose $A_1^2$-orbit is realizable, in the sense that one can identify states $\vec{s}_2$, $\vec{s}_3$, and $\vec{s}_4$ in the operational theory such that the quadruple satisfies Eqs.~\eqref{eq:rectanglesymmetryEV} and Eq.~\eqref{eq:op_equiv}.
We here show that the assumption that these states and measurements admit of a noncontextual model (and hence satisfy Eq.~\eqref{linearnc}) implies that the $\mathrm{X}$-predictability and $\mathrm{Z}$-predictability must satisfy $|\langle X\rangle| + |\langle Z\rangle| \le 1$ (Eq.~\eqref{NCURgeneral} of the main text).  We also show that this noncontextuality inequality is tight.


If $\vec{\mu}_1, \vec{\mu}_2, \vec{\mu}_3, \vec{\mu}_4$ denote the probability distributions over ontic states associated to the quadruple of operational states $\vec{s}_1, \vec{s}_2, \vec{s}_3, \vec{s}_4$, and $\vec{\xi}_{+1|X}$, $\vec{\xi}_{-1|X}$ (respectively $\vec{\xi}_{+1|Z}$, $\vec{\xi}_{-1|Z}$) are the conditional probability distributions associated to the +1 and -1 outcomes of the $X$ measurement (respectively the $Z$ measurement), then in an ontological model of these operational states and measurements, the expectation values of $X$ and $Z$ are defined as
\begin{align} \label{OMEVs}
\langle X\rangle &= (\vec{\xi}_{+1|X} -  \vec{\xi}_{-1|X})\cdot \vec{\mu}_1, \nonumber\\
\langle Z\rangle &=  (\vec{\xi}_{+1|Z} -  \vec{\xi}_{-1|Z})\cdot \vec{\mu}_1.
\end{align}
Noting that the condition of equal-predictability counterparts, Eq.~\eqref{eq:rectanglesymmetryEV}, can equivalently be written as
\begin{align}
&\mathbb{P}(+1|X,\vec{s}_1)\!=\!\mathbb{P}(-1|X,\vec{s}_2)\!=\! \mathbb{P}(-1|X,\vec{s}_3)\!=\! \mathbb{P}(+1|X,\vec{s}_4),\nonumber\\
&\mathbb{P}(+1|Z,\vec{s}_1) \!=\! \mathbb{P}(+1|Z,\vec{s}_2)\!=\! \mathbb{P}(-1|Z,\vec{s}_3)\!=\! \mathbb{P}(-1|Z,\vec{s}_4),
\end{align}\blk
it follows that the ontological model must satisfy
\begin{align}
\vec{\xi}_{+1|X}\!\cdot\! \vec{\mu}_{1} \!=\!  \vec{\xi}_{-1|X}\!\cdot\! \vec{\mu}_{2} \!=\! \vec{\xi}_{-1|X}\!\cdot\! \vec{\mu}_{3} \!=\! \vec{\xi}_{+1|X}\!\cdot\! \vec{\mu}_{4},\label{kkk2} \\ 
\vec{\xi}_{+1|Z}\!\cdot\! \vec{\mu}_{1} \!=\! \vec{\xi}_{+1|Z}\!\cdot\! \vec{\mu}_{2} \!=\! \vec{\xi}_{-1|Z}\!\cdot\! \vec{\mu}_{3} \!=\! \vec{\xi}_{-1|Z}\!\cdot\! \vec{\mu}_{4}.\label{kkk1}
\end{align}
Meanwhile, the operational equivalence condition, Eq.~\eqref{eq:op_equiv}, together with an instance of the assumption of preparation noncontextuality, Eq.~\eqref{linearnc}, implies 
\begin{equation}\label{PNCconsequence}
\frac{1}{2} \vec{\mu}_1 + \frac{1}{2} \vec{\mu}_3 = \frac{1}{2}\vec{\mu}_2+ \frac{1}{2}\vec{\mu}_4. 
\end{equation}
It is this pair of constraints on the ontological model that implies  a tradeoff relation between the $\mathrm{X}$-predictability and the $\mathrm{Z}$-predictability.

 To derive an upper bound on $|\langle X\rangle_{\vec{s}_1}| + |\langle Z\rangle_{\vec{s}_1}|$ for any state $\vec{s}_1$ satisfying the $A_1^2$-orbit-realizability condition,  it suffices to derive an upper bound on $\langle X\rangle_{\vec{s}_1} + \langle Z\rangle_{\vec{s}_1}$ (without the absolute values).  The reason is that 
 \begin{align} \label{symmetrylogic}
 |\langle X&\rangle_{\vec{s}_1}| + |\langle Z\rangle_{\vec{s}_1}|  \nonumber\\
 &= \max\Big\{
\langle X\rangle_{\vec{s}_1} + \langle Z\rangle_{\vec{s}_1}, \langle X\rangle_{\vec{s}_1} - \langle Z\rangle_{\vec{s}_1},\nonumber\\
&\quad \quad -\langle X\rangle_{\vec{s}_1} + \langle Z\rangle_{\vec{s}_1}, -\langle X\rangle_{\vec{s}_1} - \langle Z\rangle_{\vec{s}_1}\Big\}.  \nonumber\\
\end{align}
 and this, together with the condition of having equal-predictability counterparts, Eq.~\eqref{eq:rectanglesymmetryEV}, implies that
\begin{align} \label{symmetrylogic}
 |\langle X&\rangle_{\vec{s}_1}| + |\langle Z\rangle_{\vec{s}_1}|  \nonumber\\
& =  \max
\Big\{\langle X\rangle_{\vec{s}_1} + \langle Z\rangle_{\vec{s}_1}, \langle X\rangle_{\vec{s}_4} + \langle Z\rangle_{\vec{s}_4},\nonumber\\
& \quad \quad \quad \langle X\rangle_{\vec{s}_2} + \langle Z\rangle_{\vec{s}_2}, \langle X\rangle_{\vec{s}_3} +\langle Z\rangle_{\vec{s}_3}\Big\}.  \nonumber\\
&\le \max_{\vec{s}}  (\langle X\rangle_{\vec{s}} + \langle Z\rangle_{\vec{s}}),
\end{align}
where the maximization in the final expression is over all states satisfying the $A_1^2$-orbit-realizability condition, so that the final inequality is true by virtue of $\vec{s}_1$, $\vec{s}_2$, $\vec{s}_3$, and $\vec{s}_4$ being included in this set.


From this point onward, the basic structure of the argument follows the logic of Ref.~\cite{schmid2018all}.
The scenario involves just two binary-outcome measurements, so we can divide the ontic state space into four regions, corresponding to the four possible pairs of outcomes assigned to these.
Without loss of generality, therefore, we can consider only four ontic states, and the $\vec{\mu}$ and $\vec{\xi}$ vectors can consequently be taken to be vectors in a 4-dimensional real vector space.
We adopt the convention that
\begin{align} 
\vec{\xi}_{+1|X} &= (0,1,0,1), \label{respfuncX}\\
\vec{\xi}_{+1|Z} &= (1,1,0,0).\label{respfuncZ}
\end{align}
Normalization of states implies that $\vec{\xi}_{{-}1|X} \equiv \vec{u} - \vec{\xi}_{{+}1|X}$ and  $\vec{\xi}_{{-}1|Z} \equiv \vec{u} - \vec{\xi}_{{+}1|Z}$, where $\vec{u}=(1,1,1,1)$ is the unit effect. 

Note that we are here modelling the measurements using conditional probability distributions that give a deterministic outcome for each ontic state (such a response is said to be `outcome deterministic'~\cite{determinism}).
It was shown in Ref.~\cite{schmid2018all} that this can always be done without loss of generality if there are no nontrivial operational equivalences among the measurement effects, as is the case here.  The reason is as follows.  The ontic states can be taken to be the convexly extremal points in the polytope of noncontextual assignments to the set of measurement effects. If there are no nontrivial operational equivalences among the measurement effects, however, then there are no constraints arising from noncontextuality and the polytope is simply the set of all logically possible assignments to the set of measurement effects.  The convexly extremal elements of this polytope are outcome-deterministic.

Let the probability distribution over ontic states associated to the operational state $\vec{s}_1$ be parameterized as 
\begin{equation}\label{mu1}
\vec{\mu}_{1} = (a,b,c,d),
\end{equation}
with $a,b,c,d\in[0,1].$ 
By normalization of probability distributions, these four parameters are constrained by the equality
\begin{equation}\label{normalization}
a+b+c+d=1.
\end{equation}

The probability distribution over ontic states associated to operational state $\vec{s}_2$  can be parameterized without loss of generality as follows
\begin{equation}\label{mu2}
\vec{\mu}_{2} = (b + \epsilon,a-\epsilon,d - \epsilon,c+ \epsilon),
\end{equation}
where $\epsilon$ is a real parameter satisfying
\begin{align} \label{aepsilon}
&a-1 \le \epsilon \le a,\\ \label{bepsilon}
&-b \le \epsilon \le 1-b,\\ \label{cepsilon}
&-c \le \epsilon \le 1-c,\\ \label{depsilon}
&d-1 \le \epsilon \le d.
\end{align}
The proof that this parametrization is without loss of generality is as follows.  Suppose $\vec{\mu}_{2} = (b + \epsilon,a+\epsilon',d +\epsilon'',c+ \epsilon''')$, which is clearly a generic parametrization.  The fact that $\vec{s}_1$ and $\vec{s}_2$ have the same value of $\langle Z \rangle$ implies that $\vec{\mu}_1$ and $\vec{\mu}_2$ have the same sums over the first two components and the same sums over the second two components (via Eqs.~\eqref{kkk1} and \eqref{respfuncZ}), which implies that $\epsilon' = -\epsilon$ and $\epsilon'''=-\epsilon''$.  
The fact that $\vec{s}_1$ and $\vec{s}_2$ have opposite values of $\langle X \rangle$ implies that the sum of the first and third components of $\vec{\mu}_2$ must be equal to the sum of the second and fourth components of $\vec{\mu}_1$ (via Eqs.~\eqref{kkk2} and \eqref{respfuncX}), which implies that $\epsilon'' =\epsilon$.
Finally, the inequalities on $\epsilon$ follow from demanding that all of the components of $\vec{\mu}_2$ are valid probabilities.

The probability distribution over ontic states associated to operational state $\vec{s}_3$  can be parameterized without loss of generality as 
\begin{equation}\label{mu3}
\vec{\mu}_{3} = (d - \gamma,c+\gamma,b+\gamma,a-\gamma),
\end{equation}
where $\gamma$ is a  real parameter
satisfying
\begin{align}\label{aeta}
&a-1  \le \gamma \le a,\\ \label{beta}
&-b \le \gamma \le 1-b,\\ \label{ceta}
&-c \le \gamma \le 1-c,\\ \label{deta}
&d-1 \le \gamma \le d.
\end{align}
The proof that this parametrization is without loss of generality is analogous to the case of $\vec{\mu}_{2}$.

Finally, the probability distribution over ontic states associated to operational state  $\vec{s}_4$ can be parameterized without loss of generality as follows 
\begin{equation}\label{mu4}
\vec{\mu}_{4} = (c + \delta, d - \delta,a -  \delta,b + \delta ),
\end{equation}
where $\delta$ is a real parameter satisfying
\begin{align}\label{adelta}
&a-1 \le \delta \le a,\\ \label{bdelta}
&-b \le \delta \le 1-b,\\ \label{cdelta}
&-c \le \delta \le 1-c,\\ \label{ddelta}
&d-1 \le \delta \le d.
\end{align}
Again, the proof is analogous to the one above.

We start by noting that this parameterization, together with Eq.~\eqref{PNCconsequence}, implies 
\begin{align}
a+d  = b+c +(\epsilon + \gamma+ \delta).
\end{align}
Combining this constraint with the normalization condition, Eq.~\eqref{normalization}, we obtain
\begin{align}\label{abcd2a}
a+d  =\frac{1}{2}+ \frac{\epsilon + \gamma + \delta}{2}\\ \label{abcd2}
b+c = \frac{1}{2} - \frac{\epsilon + \gamma + \delta}{2} .
\end{align}


For brevity, we drop the $\vec{s}_1$ subscript from the expectation values of $X$ and $Z$.  Given Eqs~\eqref{OMEVs}, ~\eqref{respfuncX}, \eqref{respfuncZ} and \eqref{mu1}, it follows that
\begin{align}\label{ZXabcd}
&\langle X\rangle = b+ d  - a - c,\nonumber\\
&\langle Z\rangle = a+b -c-d,
\end{align}
which in turn implies that
\begin{equation} \label{XZ2bc}
\langle X\rangle + \langle Z\rangle= 2(b-c).
\end{equation}

We can now evaluate the upper bound on the latter equation, 
\begin{align}\label{XZ1c}
\langle X\rangle + \langle Z\rangle=&2\left(b-c\right)= 2\left(b+c\right)-4c \nonumber\\
= & 1-\left(\epsilon+\gamma+ \delta \right)-4c \nonumber\\ 
\le & 1+3c-4c=1-c\leq1, 
\end{align}
 where, in the second line we have used Eq.~\eqref{abcd2}, and in the third line we have used the inequalities \eqref{cepsilon}, \eqref{ceta}, and \eqref{cdelta} to minimize $\left(\epsilon+\gamma+ \delta \right)$. 
  This concludes the proof.  

The single nonlinear inequality $|\langle X\rangle| + |\langle Z\rangle| \le 1$ that we have proven can be expressed equivalently as four inequalities on linear combinations of $\langle X\rangle$ and $\langle Z\rangle$, namely,
\begin{subequations}\label{NCURs2}
\begin{align}
\langle X\rangle + \langle Z\rangle \le 1, \\
\langle X\rangle - \langle Z\rangle \le 1, \\
-\langle X\rangle + \langle Z\rangle \le 1, \\
-\langle X\rangle - \langle Z\rangle \le 1.
\end{align}
\end{subequations}

\begin{figure}[htbp] 
   \centering
   \includegraphics[width=1.9in]{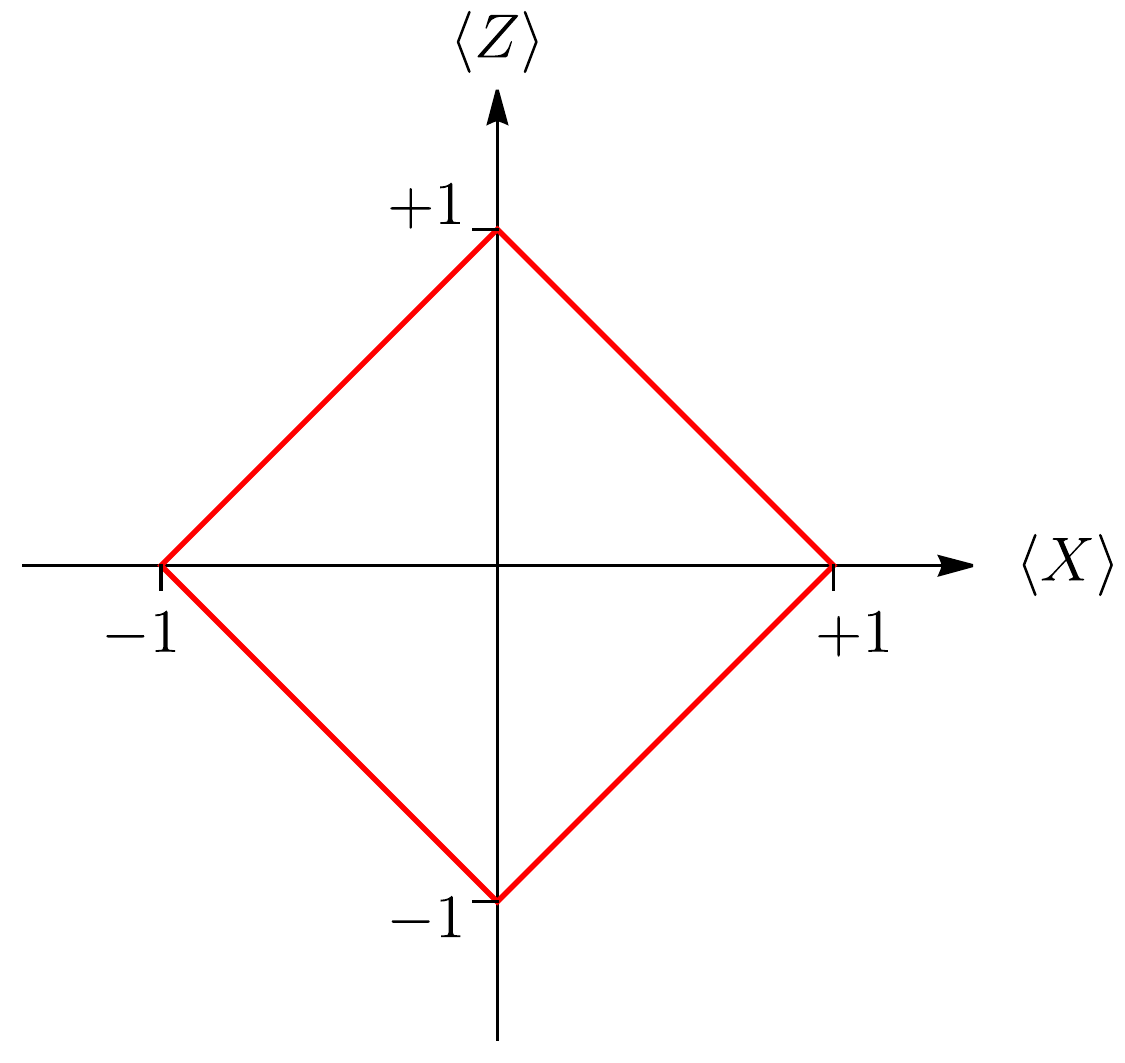}
   \caption{The noncontextually realizable expectation values of $X$ and $Z$ for states that satisfy the $A_1^2$-orbit realizability condition. }
   \label{fig:diamond}
\end{figure}

In the space of possible values of $\langle X\rangle$ and $\langle Z\rangle$, these describe a diamond, depicted in Fig.~\ref{fig:diamond}.  


As an aside, we note that {\em unlike} the absolute values of  expectation values (such as $|\langle X\rangle|$ and $|\langle Z\rangle|$), the expectation values themselves (such as $\langle X\rangle$ and $\langle Z\rangle$) are not measures of predictability. As such, the four inequalities in Eq.~\eqref{NCURs2} do not individually express constraints on predictabilities of measurements. However, one can still view each of these inequalities as a kind of {\em fine-grained} uncertainty relation, insofar as it expresses constraints on the outcome statistics of a pair of measurements achievable by a single state.\footnote{One does not need to express uncertainty relations in terms of traditional measures of unpredictability such as variance or entropy. If the probability distributions over the outcomes of two measurements are denoted by vectors ${\vec p}_{1}$ and ${\vec p}_{2}$, any tradeoff relation of the form $f({\vec p}_1,{\vec p}_2) \le C$  for some function $f$ and constant $C$ can express an uncertainty relation.  Examples of this general form of uncertainty relation, where the function $f$ is linear, have been presented, for example, in Ref.~\cite{Oppenheim2010} under the name of {\em fine-grained uncertainty relations}. }

We now turn to demonstrating that the noncontextuality inequality we have derived is tight by exhibiting an example of a noncontextual model that can achieve any point on the bounding curve.

 It is clear from the last inequality in Eq.~\eqref{XZ1c} that to saturate this upper bound, we must set $c=0$.
Setting $\epsilon=0$, $\gamma=0$ and $\delta=0$ as well, we can infer that $b=1/2$ via Eq.~\eqref{abcd2}. 
In this example, the four probability distributions over ontic states take the following simple form:
\begin{align}
\vec{\mu}_{1} &= (1/4+u,1/2,0,1/4-u),\label{concmu1c0}\\
\vec{\mu}_{2} &= (1/2,1/4+u,1/4-u,0),\label{concmu2c0}\\
\vec{\mu}_{3} &= (1/4-u,0,1/2,1/4+u),\label{concmu3c0}\\
\vec{\mu}_{4} &= (0,1/4-u,1/4+u,1/2)\label{concmu4c0},
\end{align}
where the parameter $u\in [-1/4,1/4]$ determines how $a$ and $d$ share the probability $1/2$ between them. Notice that, with this form of the $\vec{\mu}$s, the expectation values of $X$ and $Z$ take the form 
$\langle X \rangle=1-u'$ and $\langle Z \rangle=u',$ for $u'=1/2 +2u\in[0,1],$ which means that varying among all the values of $u$ (and so $u'$), one can achieve any point that saturates the inequality $\langle X \rangle + \langle Z \rangle \le 1$. 
By virtue of the $A_1^2$ symmetry of the problem, one can construct examples that saturate the other inequalities in Eq.~\eqref{NCURs2} as well. 
We conclude that one can achieve any point saturating the noncontextuality inequality of Eq.~\eqref{NCURgeneral}, establishing that the latter is tight.  
\blk

Fig. 7 depicts some examples of the four distributions of Eqs.~\eqref{concmu1c0}-\eqref{concmu4c0} for different values of $u'$, and illustrates the point on the noncontextual bound that each corresponds to.

 \begin{figure}[htbp] 
    \centering
    \includegraphics[width=.49\textwidth]{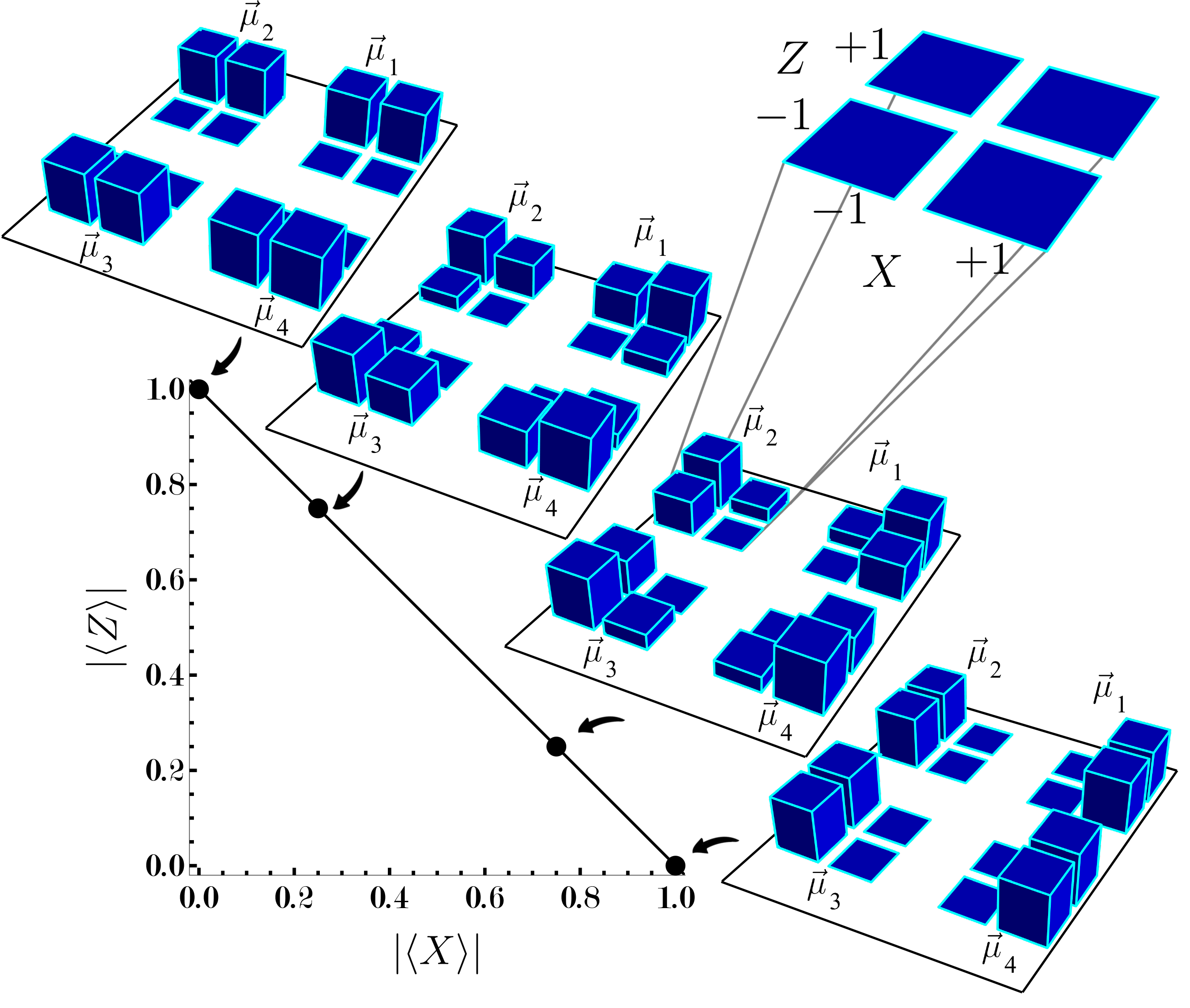} 
    \caption{Four quadruples of probability distributions over ontic states that satisfy the $A_1^2$-orbit-realizability condition and that saturate the linear tradeoff between the $\mathrm{Z}$-predictability and the $\mathrm{X}$-predictability. Note that the  condition that the state has equal-predictability counterparts is manifest in the shapes of these distributions (see the legend regarding the labelling of the ontic states in terms of the $X$ and $Z$ measurement outcomes they predict).  Similarly, the operational equivalence of the equal mixture of the two distributions on one diagonal ($\vec{\mu}_1$ and $\vec{\mu}_3$)  and the two on the opposite diagonal ($\vec{\mu}_2$ and $\vec{\mu}_4$) is also easily verified by eye.}
    \label{fig:LinearTradeoffIllustration}
 \end{figure}
 
\section{Alternative way of obtaining the noncontextuality inequalities}
\label{appendix_AlternativeProof}

The complete set of noncontextuality inequalities that hold for two binary-outcome measurements and four preparations that have the operational equivalence relation of Eq.~\eqref{eq:op_equiv} have been worked out previously in Ref.~\cite{schmid2018all}.  This is a relaxation of the problem considered here, since in Ref.~\cite{schmid2018all}, the four preparations are not required to satisfy the $A_1^2$-symmetry condition.

Defining the eight parameters $\mathbb{P}_{ts} \equiv \mathbb{P}({+}1|t,s)$, where $t$ labels the measurement setting and $s$ labels the preparation, the noncontextuality inequalities  derived in Ref.~\cite{schmid2018all} are:
\begin{subequations}\label{poly}\begin{align}
&\mathbb{P}_{12}+\mathbb{P}_{22}-\mathbb{P}_{23}-\mathbb{P}_{14} \le 1 \,,\label{eq:polyviolated1}\\ 
&\mathbb{P}_{12}+\mathbb{P}_{22}-\mathbb{P}_{13}-\mathbb{P}_{24} \le 1 \,,\\ 
&\mathbb{P}_{22}+\mathbb{P}_{13}-\mathbb{P}_{12}-\mathbb{P}_{24} \le 1 \,,\\ 
&\mathbb{P}_{12}+\mathbb{P}_{23}-\mathbb{P}_{22}-\mathbb{P}_{14} \le 1 \,,\\ 
&\mathbb{P}_{22}+\mathbb{P}_{14}-\mathbb{P}_{12}-\mathbb{P}_{23} \le 1 \,,\\ 
&\mathbb{P}_{23}+\mathbb{P}_{14}- \mathbb{P}_{12} -\mathbb{P}_{22} \le 1 \,,\\ 
&\mathbb{P}_{12}+\mathbb{P}_{24}-\mathbb{P}_{22}-\mathbb{P}_{13} \le 1 \,,\\ 
&\mathbb{P}_{13}+\mathbb{P}_{24}-\mathbb{P}_{12}-\mathbb{P}_{22} \le 1 \,\label{eq:polyviolated}.
\end{align}\end{subequations}
Note that probabilities for the $s=1$ preparation do not appear here.  This is because it was possible to eliminate these using the operational equivalence relation, Eq.~\eqref{eq:op_equiv}.   (We note also that the $\mathbb{P}_{12}$ term in the sixth inequality was mistakenly written as $\mathbb{P}_{21}$ in Ref.~\cite{schmid2018all}.)
Together with the requirement that 
$$\forall t,s:\qquad 0 \le \mathbb{P}_{ts} \le 1,$$ 
these fully characterize the polytope of noncontextually realizable correlations.

Associating the measurement setting $t=1$ to $Z$
and the setting $t=2$ to 
$X$, so that
\begin{equation}
\mathbb{P}_{1s} = \frac{\langle Z \rangle_{s} +1}{2},\qquad
\mathbb{P}_{2s} = \frac{\langle X \rangle_{s} +1}{2},
\end{equation}
we can simplify the noncontextuality inequalities of Eq.~\eqref{poly} using the $A_1^2$-symmetry condition (Eq.~\eqref{eq:rectanglesymmetryEV}) and then re-express them in terms of $\langle Z\rangle \equiv \langle Z \rangle_{s}$ and $\langle X\rangle \equiv \langle X \rangle_{s}$. 
Doing so, one finds that the eight linear inequalities of Eq.~\eqref{poly} reduce to the four linear inequalities of Eq.~\eqref{NCURs2}, and consequently to the single nonlinear inequality of Eq.~\eqref{NCURgeneral}.

\section{Extension of results to the case of three measurements} 
\label{Appendix_3Dcase}

It is useful to consider the generalization of our results to {\em three} measurements, corresponding in the qubit theory to $X$, $Y$ and $Z$ Pauli observables.  We have already noted that in the qubit theory, there is a nontrivial tradeoff among these, given in Eq.~\eqref{quantumXYZUR} and termed the quantum $\mathrm{XYZ}$-uncertainty relation.  The $\mathrm{XYZ}$-uncertainty relations of our other four foil theories are as follows:
\begin{align}
\textrm{qubit stabilizer:}& \;\;  |\langle X\rangle| + |\langle Y\rangle| + |\langle Z\rangle|  \le 1,\label{stabUR3d}\\
\textrm{$\eta$-depolarized qubit:}&\;\; \langle X\rangle^2 + \langle Y\rangle^2+ \langle Z\rangle^2 \le (1- \eta)^2, \label{etaUR3d}\\
\textrm{gbit:}& \;\;  |\langle X\rangle| \le 1, |\langle Y\rangle| \le 1, |\langle Z\rangle| \le 1\label{gbitUR3d}\\
 \textrm{simplicial:}& \;\; |\langle X\rangle| \le 1, |\langle Y\rangle| \le 1, |\langle Z\rangle| \le 1.\label{simplicialUR3d}
\end{align}
These are depicted in Fig.~\ref{Tradeoffs}(b) alongside the quantum $\mathrm{XYZ}$-uncertainty relation. Note that the relations for the gbit and simplicial theory again describe a lack of any nontrivial tradeoff, as $X$, $Y$ and $Z$ can all be made perfectly predictable by a single state. 

\begin{figure}[htb!] 
	\centering
        \includegraphics[width=2.5in]{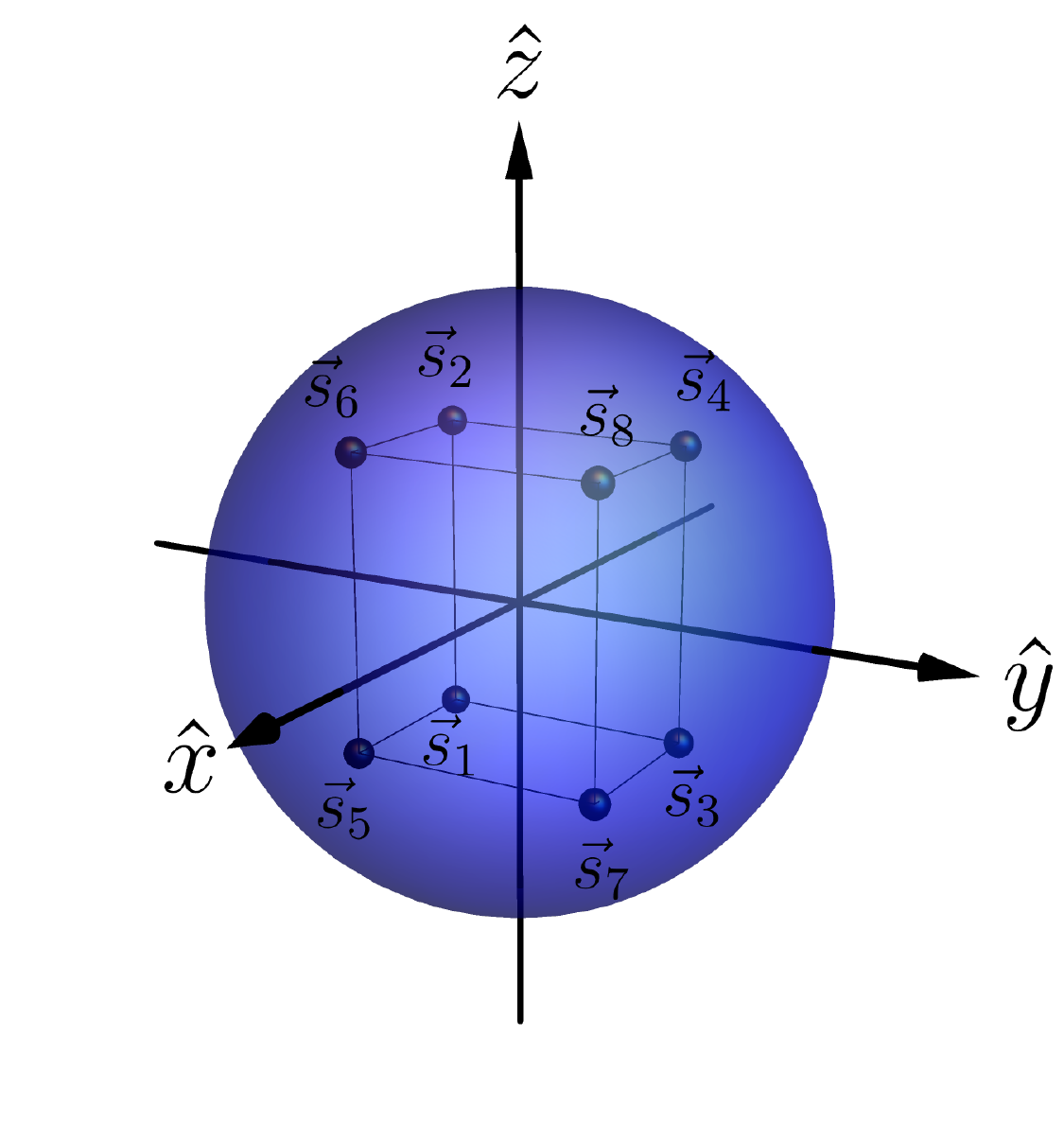}
	\caption{  Depiction of how an arbitrary state $\vec{s}_1$ in qubit quantum theory is part of an octuplet of states that satisfy the $A_1^3$-orbit-realizability condition.  }
	\label{fig:labeling}
\end{figure}

Following logic analogous to that in the main text, we now study the consequences of noncontextuality for the tradeoff in predictabilities of these three measurements.
We begin (as in the main text) by considering the quantum case, and noting that for each state of a qubit, one can find seven other states with the same $\mathrm{X}$-predictability, $\mathrm{Y}$-predictability and $\mathrm{Z}$-predictability and where the eight states cover the $2^3$ possible values of $\langle X\rangle, \langle Y\rangle$, and $\langle Z\rangle$.  Denoting the real-valued vectors representing these eight states (i.e., their Bloch representations) by $\vec{s}_1, \dots, \vec{s}_8$, this condition can be expressed as
\begin{align}\label{equalpredictabilityXYZ}
& \langle X\rangle_{\vec{s}_1}= \langle X\rangle_{\vec{s}_2}= \langle X\rangle_{\vec{s}_3}=\langle X\rangle_{\vec{s}_4} \\
&=- \langle X\rangle_{\vec{s}_5} = - \langle X\rangle_{\vec{s}_6} = -\langle X\rangle_{\vec{s}_7}= -\langle X\rangle_{\vec{s}_8},\nonumber\\
& \langle Y\rangle_{\vec{s}_1}= \langle Y\rangle_{\vec{s}_2}= \langle Y\rangle_{\vec{s}_5}=\langle Y\rangle_{\vec{s}_6} \nonumber\\
&=- \langle Y\rangle_{\vec{s}_3} = - \langle Y\rangle_{\vec{s}_4} = -\langle Y\rangle_{\vec{s}_7}=-\langle Y\rangle_{\vec{s}_8},\nonumber\\
& \langle Z\rangle_{\vec{s}_1}= \langle Z\rangle_{\vec{s}_3}= \langle Z\rangle_{\vec{s}_5}=\langle Z\rangle_{\vec{s}_7} \nonumber\\
&=- \langle Z\rangle_{\vec{s}_2} = - \langle Z\rangle_{\vec{s}_4} = -\langle Z\rangle_{\vec{s}_6}= -\langle Z\rangle_{\vec{s}_8}.\nonumber
\end{align}
Such sets of eight states form a rectangular prism, as depicted in Fig.~\ref{fig:labeling}. As they can all be generated by the orbit of the first state, $\vec{s}_1$, under the symmetry group of a rectangular prism, i.e., the Coxeter group  $A_1^3$, we refer to this condition on the state as the {\em $A_1^3$-orbit-realizability condition}.

Any eight such states necessarily satisfy the following operational equivalence relations: 
\begin{align}\label{OEsXYZ}
\frac{1}{2} \vec{s}_8 + \frac{1}{2}\vec{s}_{1} &= \frac{1}{4} \vec{s}_1 + \frac{1}{4} \vec{s}_4 + \frac{1}{4} \vec{s}_6 +\frac{1}{4} \vec{s}_7\nonumber\\
\frac{1}{2} \vec{s}_5 + \frac{1}{2}\vec{s}_{4} &= \frac{1}{4} \vec{s}_1 + \frac{1}{4} \vec{s}_4 + \frac{1}{4} \vec{s}_6 +\frac{1}{4} \vec{s}_7\nonumber\\
\frac{1}{2} \vec{s}_3 + \frac{1}{2}\vec{s}_{6} &= \frac{1}{4} \vec{s}_1 + \frac{1}{4} \vec{s}_4 + \frac{1}{4} \vec{s}_6 +\frac{1}{4} \vec{s}_7\nonumber\\
\frac{1}{2} \vec{s}_2 + \frac{1}{2}\vec{s}_{7} &= \frac{1}{4} \vec{s}_1 + \frac{1}{4} \vec{s}_4 + \frac{1}{4} \vec{s}_6 +\frac{1}{4} \vec{s}_7.
\end{align}
It is easy to verify geometrically that the four operational equivalence relations  in Eq.~\eqref{OEsXYZ} hold for these eight states, since each simply describes two different ensembles of states for which the ensemble-average is the completely mixed state. Furthermore, one can see that there are no further operational equivalences\footnote{More precisely, there are no further operational equivalences in the case where the eight states are all distinct. 
} that are logically independent of these. This is because four of the eight vectors under consideration are linearly independent (e.g., $\vec{s}_1$, $\vec{s}_4$, $\vec{s}_6$, and $\vec{s}_7$, which form the vertices of a regular tetrahedron), and the four operational equivalence relations above simply express the decomposition of the remaining four vectors in terms of these four.

Note that these operational equivalences hold for {\em any} eight states that are equal-predictability counterparts of one another in the sense of Eq.~\eqref{equalpredictabilityXYZ}. This is a point of contrast with the situation for a pair of measurements considered in the main text, where the operational equivalence relation of Eq.~\eqref{eq:op_equiv} did not follow from  the mere promise that four states were equal-predictability counterparts of one another (Eq.~\eqref{eq:rectanglesymmetryEV}), but needed to be imposed as an additional constraint. Thus, while $A_1^2$-orbit-realizability was defined as the conjunction of Eq.~\eqref{eq:rectanglesymmetryEV} and Eq.~\eqref{eq:op_equiv}, $A_1^3$-orbit-realizability is defined simply as the condition of Eq.~\eqref{equalpredictabilityXYZ}.

 As noted earlier, states can be represented by real-valued vectors not just in quantum theory, but in any operational theory.  Consequently, the $A_1^3$-orbit-realizability condition of Eq.~\eqref{equalpredictabilityXYZ} can be articulated as a condition on a state in an arbitrary operational theory (relative to any triple of measurements therein).  

We can now state the three-measurement analogue of our main result.  In any operational theory, if one can find a triple of measurements (which we denote by $X$, $Y$, and $Z$) and a state that satisfies the $A_1^3$-orbit-realizability condition relative to this triple, 
 then noncontextuality implies a nontrivial constraint on the predictabilities $|\langle X\rangle|$, $|\langle Y\rangle|$ and $|\langle Z\rangle|$ for that state, namely, that they satisfy:
\begin{align}\label{NCURgeneralappendix}
|\langle X\rangle| +|\langle Y\rangle| + |\langle Z\rangle| \le 1.
\end{align}
 This noncontextuality inequality is the generalization (from two to three measurements) of  the noncontextuality inequality of Eq.~\eqref{NCURgeneral} from the main text. The proof is also exactly analogous, following the logic of Appendix~\ref{appendix_MainProof}. As the quantifier elimination problem becomes much more difficult than in the case of two measurements, we do not provide an analytic proof here. It is straightforward to verify the result using computational algebra.  One can also reduce the problem  to a linear program in the manner described in Ref.~\cite{schmid2018all} and solve the latter computationally.

 

Whether an operational theory has $A_1^3$-symmetry or not, Eq.~\eqref{NCURgeneralappendix} constrains the tradeoff between $\mathrm{X}$-predictability, $\mathrm{Y}$-predictability, and $\mathrm{Z}$-predictability for any state within the theory that satisfies the $A_1^3$-orbit-realizability condition.  Consequently, if the theory contains one or more such states that {\em violate} the inequality, this is a proof of the failure of that theory to admit of a noncontextual ontological model. 
For operational theories that {\em do} have $A_1^3$-symmetry, Eq.~\eqref{NCURgeneralappendix} has further significance.  Because in such theories {\em all states} satisfy the $A_1^3$-orbit-realizability condition, our bound is a universal constraint on the predictability tradeoff within such theories, that is, it is a constraint on
{\em the form of the $\mathrm{XYZ}$-uncertainty relation} within such theories. 

The noncontextual bound (Eq.~\eqref{NCURgeneralappendix}) is compared to the $\mathrm{XYZ}$-uncertainty relation for a qubit (Eq.~\eqref{quantumXYZUR}) in Fig.~\ref{Tradeoffs}(b), where it is readily seen that there can be quantum violations of the bound.   Indeed, only when $|\langle X\rangle|=1$ or $|\langle Y\rangle|=1$ or $|\langle Z\rangle|=1$ does the noncontextual bound intersect the quantum $\mathrm{XYZ}$-uncertainty relation. The maximum quantum violation is achieved when $|\langle X\rangle|= |\langle Y\rangle|= |\langle Z\rangle|= \frac{1}{\sqrt{3}}$ and corresponds to 
$|\langle X\rangle| + |\langle Y\rangle| + |\langle Z\rangle|= \sqrt{3} \simeq 1.732 $.
Note that this is a larger relative violation than is possible for the inequality based on two Pauli observables, Eq.~\eqref{NCURgeneral}, for which quantum theory achieves $|\langle X\rangle| + |\langle Z\rangle| = \sqrt{2} \simeq 1.414$.   

We now compare this noncontextual bound with the $\mathrm{XYZ}$-uncertainty relation of the three foil theories that are in the $A_1^3$-symmetry class.  The $\mathrm{XYZ}$-uncertainty relation for the $\eta$-depolarized qubit theory, Eq.~\eqref{etaUR3d}, satisfies the noncontextual bound if $\eta \le 1- \frac{1}{\sqrt{3}}\simeq 0.423$.  The uncertainty relation for the qubit stabilizer theory, Eq.~\eqref{stabUR3d}, has exactly the same form as Eq.~\eqref{NCURgeneralappendix} and therefore precisely saturates the noncontextual bound.   Finally, the uncertainty relation for the gbit theory,  Eq.~\eqref{gbitUR3d}, yields the algebraic maximum possible violation of the noncontextual bound, namely, $|\langle X\rangle| + |\langle Y\rangle| + |\langle Z\rangle| = 3$.   

By contrast, because the simplicial theory is {\em not} in the $A_1^3$-symmetry class, our result does not constrain the form of its $\mathrm{XYZ}$-uncertainty relation. Therefore, although the relation for the simplicial theory  (Eq.~\eqref{simplicialUR3d}) is equivalent to that of the gbit theory (Eq.~\eqref{gbitUR3d}) and thus can violate the inequality of Eq.~\eqref{NCboundXYZ}, the only states in the theory that achieve this violation (for example, the vertices of the simplex) do not satisfy $A_1^3$-orbit realizability and therefore the bound is not applicable to them. Meanwhile, all of the states that {\em do} satisfy the $A_1^3$-orbit realizability---those inside of the embedded octahedron---satisfy the bound.  In short, contextuality is not witnessed in the case of the simplicial theory, consistent with the fact that the latter admits of a noncontextual model. 

\begin{figure}[htbp] 
   \centering
\includegraphics[width=2.5in]{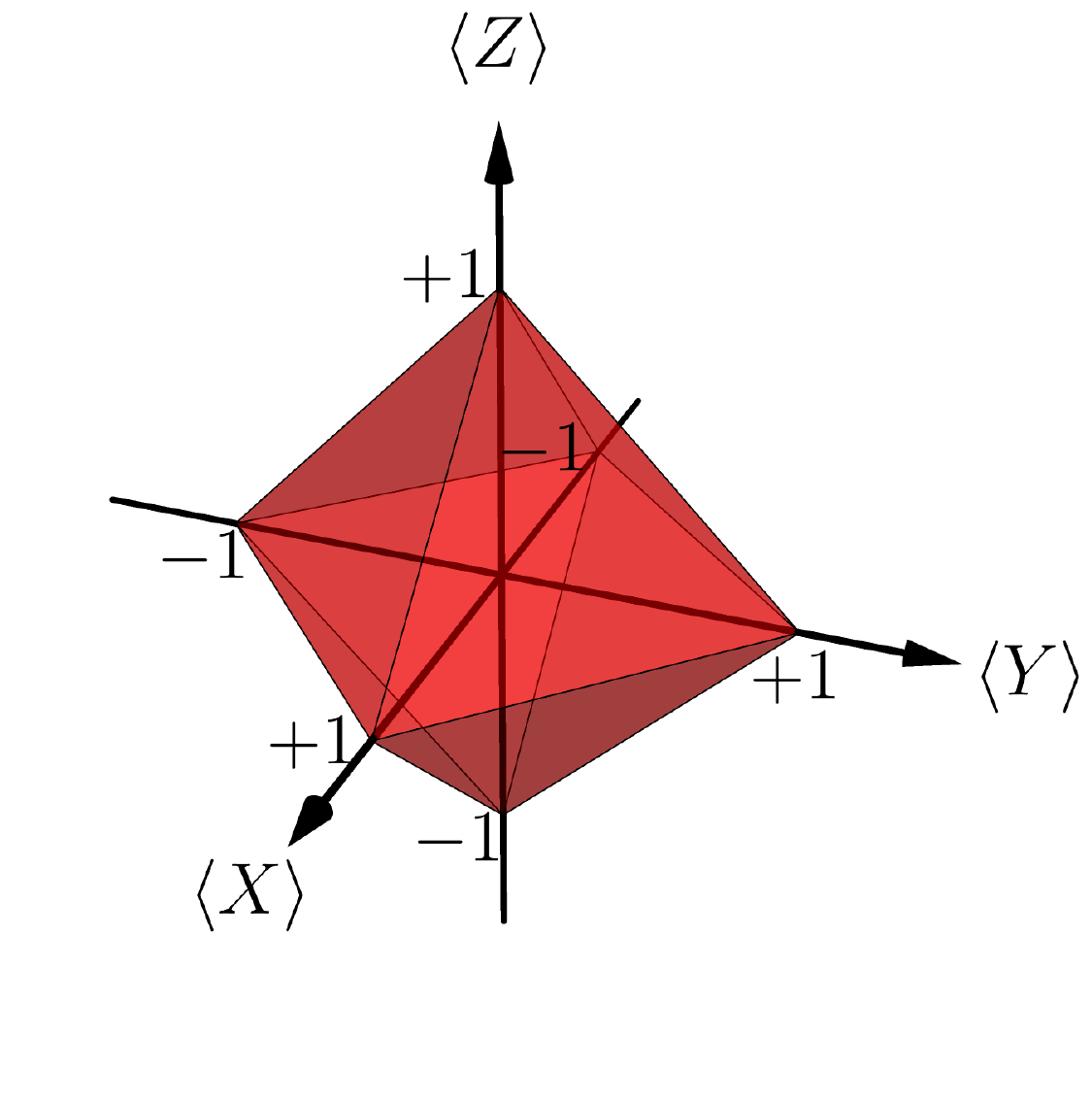}
\caption{The noncontextually realizable expectation values of $X$, $Y$, and $Z$ for states that satisfy the $A_1^3$-orbit realizability condition.}
\label{fig:octahedron}
\end{figure}

  Finally, note that, just as in the case of the two measurements, the single nonlinear noncontextuality inequality of Eq.~\eqref{NCURgeneralappendix} can also be expressed as a set of linear inequalities, namely,  \begin{subequations}\label{NCURsXYZ}
 \begin{align}
\langle X\rangle +\langle Y\rangle + \langle Z\rangle \le 1,\nonumber\\
\langle X\rangle +\langle Y\rangle - \langle Z\rangle \le 1,\nonumber\\
\langle X\rangle -\langle Y\rangle + \langle Z\rangle \le 1,\nonumber\\
\langle X\rangle -\langle Y\rangle - \langle Z\rangle \le 1,\nonumber\\
{-}\langle X\rangle +\langle Y\rangle + \langle Z\rangle \le 1,\nonumber\\
{-}\langle X\rangle +\langle Y\rangle - \langle Z\rangle \le 1,\nonumber\\
{-}\langle X\rangle -\langle Y\rangle + \langle Z\rangle \le 1,\nonumber\\
{-}\langle X\rangle -\langle Y\rangle - \langle Z\rangle \le 1.\nonumber
\end{align}
\end{subequations}
Again, these inequalities can be considered as bounds on fine-grained uncertainty relations.
 The space of values of $\langle X\rangle$, $\langle Y\rangle$, and $\langle Z\rangle$ consistent with these inequalities constitutes an octahedron, depicted in Fig.~\ref{fig:octahedron}, with each inequality describing one of the facets.

It is worth noting that an operational theory with an octahedral state space has previously been derived axiomatically by starting from a classical theory and assuming an epistemic restriction~\cite{Spekkens2007,chiribella2016quantum}.   The results here demonstrate that if one starts with the landscape of possible operational theories for a system with a real-valued vector representation of dimension 4, and one takes as axioms that the state space has $A_1^3$-symmetry and that one can realize all states consistent with noncontextuality, then one can also derive the octahedral state space.  This suggests that it might be worthwhile to try to better understand the starting point of epistemically restricted statistical theories~\cite{Spekkens2007,Bartlett2012,chiribella2016quantum,ToyFieldTheory}---specifically, what is assumed about the form of the ontic state space and the form of the epistemic restriction---from the perspective of what symmetry properties are encoded therein.

\section{Uncertainty relations for a qubit}
\label{Appendix_UR}

We here describe the strongest 
uncertainty relation for a qubit and we demonstrate various forms in which it can be expressed. 

We denote the $\pm1$ eigenstates of Pauli observables $X$, $Y$ and $Z$ as $\left\vert\pm x\right\rangle,\left\vert\pm y\right\rangle, \left\vert\pm z\right\rangle$, respectively. 
Letting $p_{x}=\mathrm{Tr} \left(  \rho\left\vert x\right\rangle \left\langle x\right\vert \right),$ $p_{y}=\mathrm{Tr}
\left(  \rho\left\vert y\right\rangle \left\langle y\right\vert \right)$, $p_{z}=\mathrm{Tr}
\left(  \rho\left\vert z\right\rangle \left\langle z\right\vert \right)$, the set of valid quantum states lie inside the Bloch ball,
corresponding to the constraint
\begin{equation}\label{Blochball1}
\left(  p_{x}-\frac{1}{2}\right)  ^{2}+\left(  p_{y}-\frac{1}{2}\right)
^{2}+\left(  p_{z}-\frac{1}{2}\right)  ^{2}\leq\frac{1}{4}.
\end{equation}

One can also express the constraint defining the Bloch ball in terms of the predictabilities defined in the text, that is, the absolute values of the expectation values of $X,Y$ and
$Z$. Recalling that
 \begin{align*}
\left\langle X\right\rangle  &  =\mathrm{Tr}\left[  \rho(\left\vert
+x\right\rangle \left\langle +x\right\vert -\left\vert -x\right\rangle
\left\langle -x\right\vert \right)] \\
&  =2p_{x}-1,
\end{align*}
so that the $\mathrm{X}$-predictability is
 \begin{align*}
|\langle X\rangle|  &  = |2p_{x}-1|,
\end{align*}
and similarly for $Y$ and $Z$, Eq.~\eqref{Blochball1} can be also expressed as 
\begin{equation}\label{Blochball2}
\left\langle X\right\rangle ^{2}+\left\langle Y\right\rangle
^{2}+\left\langle Z\right\rangle ^{2}\leq1,
\end{equation}
where we do not bother to write the absolute value explicitly when the quantity is squared.

We can also rewrite this in terms of standard deviations or, more precisely, variances. The variance of $X$, $\Delta X^{2}$, is related to the expectation value by
\begin{align*}
\Delta X^{2}  &  =\left\langle X^{2}\right\rangle
-\left\langle X\right\rangle ^{2}\\
&  =\left\langle \mathbb{I} \right\rangle -\left\langle X\right\rangle ^{2}\\
&  =1-\left\langle X\right\rangle ^{2},
\end{align*}
where we have made use of the identity $X^{2}=\mathbb{I}$.  The analogous relations hold for $Y$ and $Z$.
Therefore, Eqs.~\eqref{Blochball1} and \eqref{Blochball2} 
 can also be written as
\begin{equation}\label{Blochball3}
\Delta X^{2}+\Delta Y^{2}+\Delta Z^{2}\geq2.
\end{equation}
To our knowledge, this form of the uncertainty relation first appears in Ref.~\cite{Luis2003}, which builds the work of Ref.~\cite{Larsen1990}.

We mention one final form of this uncertainty relation.
Ref.~\cite{Luis2003} defines the {\em certainty} for $X$, denoted $C_{x}$, to be
\begin{align*}
C_{x}^{2}  &\equiv p_{x}^{2}+(1-p_{x})^{2},
\end{align*}
so that
\begin{align*}
\left(  p_{x}-\frac{1}{2}\right)  ^{2}  
&  =\frac{1}{2}C_{x}^{2}-\frac{1}{4}
\end{align*}
and similarly for $Y$ and $Z$.  
It follows that Eqs.~\eqref{Blochball1},\eqref{Blochball2},\eqref{Blochball3} can also be expressed as
\begin{equation}\label{Blochball4}
C_{x}^{2}+C_{y}^{2}+C_{z}^{2}\leq 2.
\end{equation}
The terminology introduced for $C_{x}$, $C_y$ and $C_z$ stems from the fact that they measure the degree of certainty about the outcomes of the Pauli measurements,  rather than  the degree of uncertainty about these.  Eq.~\eqref{Blochball4} is termed a ``certainty relation'' in Ref.~\cite{Luis2003}.

In this article, we have preferred to use the absolute values of the expectation values of Pauli observables, which are related to the certainties by
\begin{equation}
|\langle X\rangle| = \frac{1}{\sqrt{2}} C_x,
\end{equation}
and similarly for $Y$ and $Z$, and to refer to these as measures of {\em predictability}.
(Both the certainty and the predictability measures vary inversely to the standard deviations $\Delta X$, $\Delta Y$, and $\Delta Z$, which are measures of uncertainty.)
Since $\langle X\rangle^2  = |\langle X\rangle|^2$, and similarly for $Y$ and $Z$, Eq.~\eqref{Blochball2} can be understood as a tradeoff relation between the predictabilities.

Insofar as Eqs.~\eqref{Blochball1}-\eqref{Blochball4} all describe the Bloch ball, they are the strongest possible uncertainty relation for a qubit.  To our knowledge, this uncertainty relation was first presented in Ref.~\cite{Larsen1990}.  We refer to it here as the $\mathrm{XYZ}$-uncertainty relation.

Notice that the uncertainty relations above involve the three observables $X,Y, Z$. However, they imply 
the uncertainty relations for two observables only, like in the case of the uncertainty relation Eq.~\eqref{QUR0} that we consider in the main text. 

Starting from  Eq.~\eqref{Blochball1},
the positivity of $\left(  p_{y}-\frac{1}{2}\right)  ^{2}$ implies that 
\begin{align*}
\left(  p_{x}-\frac{1}{2}\right)  ^{2}+\left(  p_{z}-\frac{1}{2}\right)  ^{2}
&  \leq\frac{1}{4}.
\end{align*}
This relation is easily shown to be equivalent to
\begin{align*}
\left\langle X\right\rangle ^{2}+\left\langle Z\right\rangle
^{2}  &  \leq 1,
\end{align*}
and to
\begin{align*}
\Delta X^{2}+\Delta Z^{2}  &  \geq1,
\end{align*}
and to%
\begin{align*}
C_{x}^{2}+C_{z}^{2}  &  \leq\frac{3}{2}.
\end{align*}
The second and third forms are the ones appearing in the main text of the article (see the discussion around Eq.~\eqref{QUR0}).
To our knowledge, this uncertainty relation for $X$ and $Z$ first appeared in Ref.~\cite{Luis2003}, in the fourth form just described, where it was derived in much the same way as we have done here, starting from the 
 uncertainty relation for three Pauli observables.

Note that the $\mathrm{XYZ}$-uncertainty relation can also be conceptualized as  a {\em state-dependent} $\mathrm{ZX}$-uncertainty relation, namely,
\begin{align}\label{SDUR}
\Delta X^{2}+\Delta Z^{2}  &  \geq 2 - \Delta Y^{2}.
\end{align}
To get back to the state-{\em independent} $\mathrm{ZX}$-uncertainty relation $\Delta X^{2}+\Delta Z^{2}  \geq 1$, it suffices to minimize the right-hand side under a variation over the state.  This occurs when the state is in the plane of the Bloch sphere wherein $\left\langle Y\right\rangle = 0$, so that $\Delta Y^{2}=1$.  Only in the $\left\langle Y\right\rangle = 0$ plane can one saturate the state-independent $\mathrm{ZX}$-uncertainty relation.  Outside this plane, one has a tighter bound.   Indeed, in the extreme case of an eigenstate of $Y$, we have $\left\langle Y\right\rangle ^{2}=1$  and hence $\Delta Y^{2}=0$, so that $\left\langle X\right\rangle ^{2}+\left\langle Z\right\rangle^{2}  \geq 2$.  Because each of the terms on the left-hand side are bounded above by 1, saturating this inequality requires that they both be equal to 1.  This simply captures the fact that for an eigenstate of $Y$, it is indeed the case that $\left\langle X\right\rangle = 0$ and $\left\langle Z\right\rangle=0$, and hence that $\Delta  X^{2}=1$ and $\Delta Z ^{2}=1$.

Suppose one starts with the usual state-dependent uncertainty relation for $X$ and $Z$, which is of the form
\begin{align}
\Delta X^{2} \Delta Z^{2}  &\geq  \langle Y \rangle^2.
\end{align}
(This is the form one generally 
encounters in the textbooks.)
To get to an uncertainty relation that is state-independent, it suffices to minimize the right-hand side under a variation over the state.  Doing so, we obtain $\Delta X^{2} \Delta Z^{2}  \geq 0$.  However, given that $\Delta X^{2}$ and $\Delta Z^{2}$ are bounded below by 0, this inequality is trivial.  This problem has been pointed out by many authors~\cite{Deutsch1983, Maccone2014}.    By contrast, the state-dependent uncertainty relation of Eq.~\eqref{SDUR}, which involves a {\em sum} rather than a {\em product} of the standard deviations {\em does} yield a nontrivial state-independent uncertainty relation.  

Using the {\em product} of standard deviations was no doubt inspired by the original formulation of uncertainty relations for position and momentum observables by Heisenberg, Kennard, and Robertson \cite{Heisenberg1925,Kennard1927,Robertson1929}.  Many other ways of expressing the predictability tradeoffs that quantum theory implies have been studied in the literature on the subject~\cite{Hirschman1957,Beckner1975,Bialynicki1975,Deutsch1983,Maassen1988,Uffink1990,Oppenheim2010,Friedland2013,Puchala2013,Coles2017}, and many of these have been proposed specifically to solve the problem of the triviality of the 
 implications of the usual form.  Note, however, that because the $\mathrm{XYZ}$-uncertainty relation we have described above (Eqs.~\eqref{Blochball1}-\eqref{Blochball4} are the different forms of it) characterizes the Bloch ball completely, no other uncertainty relation for a qubit can express a more stringent constraint on the probability distributions over outcomes of $X$, $Y$ and $Z$ measurements.

\end{document}